\documentclass[aps,prd,twocolumn,showpacs,floatfix,preprintnumbers,amsfont,amsmath,amssymb,nofootinbib, superscriptaddress]{revtex4-1}

\usepackage{color}
\usepackage{hyperref}
\usepackage[normalem]{ulem}
\usepackage{lipsum}
\usepackage{url}
\usepackage{float}
\usepackage{ctable}
\usepackage{graphicx}
\usepackage[font=small, justification=raggedright, format=plain
  ]{caption} % 'format=plain' avoids hanging indentation
\usepackage{subcaption}
\usepackage{mathtools}
\usepackage{bm}
\usepackage{siunitx}
\usepackage{bbm}

\newcommand{\sk}[1]{}

\newcommand{\be}{\begin{equation}}
\newcommand{\ee}{\end{equation}}
\newcommand{\ba}{\begin{eqnarray}}
\newcommand{\ea}{\end{eqnarray}}

\newcommand\ddfrac[2]{\frac{\displaystyle #1}{\displaystyle #2}}

\newcommand{\pr}{\prime}

\newcommand{\intr}{\int_{-\infty}^{\infty}}
\newcommand{\llangle}{\left\langle}
\newcommand{\rrangle}{\right\rangle}
\newcommand{\llvert}{\left\vert}
\newcommand{\rrvert}{\right\vert}
\newcommand{\wt}[1]{\widetilde{#1}}
\newcommand{\conv}{\circledast}
\newcommand{\rev}{\overleftarrow}
\newcommand{\lt}{\left(}
\newcommand{\rt}{\right)}

\allowdisplaybreaks
\begin{document}

%\preprint{AAPM/123-QED}

% \title{Detecting Gravitational Waves on Data with Varying Power Spectral Density and Transient Artifacts}% Force line breaks with \\
\title{Detecting Gravitational Waves in Data with Non-Gaussian Noise}% Force line breaks with \\

\author{Barak Zackay}
\email{bzackay@ias.edu}
\affiliation{\mbox{School of Natural Sciences, Institute for Advanced Study, 1 Einstein Drive, Princeton, NJ 08540, USA}}
\author{Tejaswi Venumadhav}
% \email{tejaswi@ias.edu}
\affiliation{\mbox{School of Natural Sciences, Institute for Advanced Study, 1 Einstein Drive, Princeton, NJ 08540, USA}}
\author{Javier Roulet}
%\email{jroulet@princeton.edu}
\affiliation{\mbox{Department of Physics, Princeton University, Princeton, NJ, 08540, USA}}
\author{Liang Dai}
%\email{ldai@ias.edu}
\affiliation{\mbox{School of Natural Sciences, Institute for Advanced Study, 1 Einstein Drive, Princeton, NJ 08540, USA}}
\author{Matias Zaldarriaga}
%\email{matiasz@ias.edu}
\affiliation{\mbox{School of Natural Sciences, Institute for Advanced Study, 1 Einstein Drive, Princeton, NJ 08540, USA}}

\date{\today}% It is always \today, today,
             
%\keyword{gravitational wave detection; data analysis;}

%%%%%%%%%%%%%%%%%%%%%%%%%%%%%%%%%%%%%%%%%%%%%%%%%%%%%%%%%%%%%%%%%%%%%%%
\begin{abstract}

Searches for gravitational waves crucially depend on exact signal processing  of noisy strain data from gravitational wave detectors, which are known to exhibit significant non-Gaussian behavior.
In this paper, we study two distinct non-Gaussian effects in the LIGO/Virgo data which reduce the sensitivity of searches: first, variations in the noise power spectral density (PSD) on timescales of more than a few seconds; and second, loud and abrupt transient `glitches' of terrestrial or instrumental origin.
We derive a simple procedure to correct, at first order, the effect of the variation in the PSD on the search background.
Given the knowledge of the existence of localized glitches in particular segments of data, we also develop a method to insulate statistical inference from these glitches, so as to cleanly excise them without affecting the search background in neighboring seconds.
We show the importance of applying these methods on the publicly available LIGO data, and measure an increase in the detection volume of at least $15\%$ from the PSD-drift correction alone, due to the improved background distribution. 
\end{abstract}
%%%%%%%%%%%%%%%%%%%%%%%%%%%%%%%%%%%%%%%%%%%%%%%%%%%%%%%%%%%%%%%%%%%%%%%

%Use showkeys class option if keyword
\maketitle

%%%%%%%%%%%%%%%%%%%%%%%%%%%%%%%%%%%%%%%%%%%%%%%%%%%%%%%%%%%%%%%%%%%%%%%
\section{Introduction}
%%%%%%%%%%%%%%%%%%%%%%%%%%%%%%%%%%%%%%%%%%%%%%%%%%%%%%%%%%%%%%%%%%%%%%%

The first detection of Gravitational Waves (GWs) from merging compact binaries in LIGO data opened up the opportunity to study a new family of astrophysical sources.
Multiple sources have been detected in the first and second observing runs of Advanced LIGO and Virgo (O1 and O2) \cite{GW150914, GW151226, O1catalog, GW170104, GW170608, GW170817, LIGOScientific:2018mvr, GW151216, O2BBHs}. %the focus has shifted to characterizing the astrophysical population of merging binary neutron stars and black holes.  % \jr{People not doing populations might get annoyed at that statement?}
State-of-the-art searches \cite{gstlal,PYCBCPipeline,pipelinepaper} suffer from reduced sensitivity due to non-Gaussian effects in the strain data \cite{DetChar}, which results in fewer detected sources. It is of paramount importance to understand and correct for these non-Gaussian effects in order to maximize the yield of the existing data.

We distinguish two types of non-Gaussian behaviour, both of which contribute to the tail distribution of triggers obtained by matched filtering with a template bank.
The first type is due to the changing power spectral density (PSD) of the noise that constitutes the background for the GW search. As we show, changes in the PSD on short timescales (even as short as \SI{10}{\second}) are important to capture in order to suppress over-production of candidate signals with high signal-to-noise ratio (SNR).
The second type of non-Gaussian behavior is abrupt noise transients (`glitches' on sub-second timescales) that are caused by either environmental disturbance or instrumental malfunction. The origin of many of these glitches is not yet understood~\cite{DetChar}.

In this paper, we treat the problem of detecting putative GW events in the presence of these systematic effects. 
Firstly, we propose a simple practical solution that corrects the effect of a varying PSD on short timescales, to first order in the change in the PSD. 
Secondly, we present a method to isolate the detection statistic for compact binaries from abrupt noise transients at given times and of given lengths.

Section \ref{sec:WrongPSD} is devoted to the effect of slow variations in the PSD.
We begin by reviewing matched filtering, and the process of PSD estimation, in stationary Gaussian noise.
%and the requirements that drive choices of parameters. We explain how measurement errors in the PSD lead to sensitivity losses
We demonstrate the magnitude and timescale of the PSD variations in the LIGO data, and derive how these variations cause a loss in sensitivity for searches.
We then propose a practical and simple way to cancel the first order loss in sensitivity due to the mis-estimation of the local PSD.
Essentially, this solution is to divide all computed matched filtering overlaps by their locally estimated standard deviation $\sigma_z(t)$, given by:
\begin{align}
    \sigma^2_z(t) = \frac{1}{N_{\rm a}}\sum_{t'=t-N_{\rm a}\Delta t/2}^{t+-N_{\rm a}\Delta t/2}{|z(t')|^2}\,,
\end{align}
where $z(t)$ are the matched filtering overlaps, defined in Section \ref{sec:PSDEstimation}, and $N_{\rm a}$ is the number of scores used to estimate the average.
Finally, we use GW triggers on the entirety of the publicly available O2 LIGO data \cite{gwosc_url, GWOSC} to explicitly demonstrate the dramatic effects of PSD changes on the tail distribution of matched filtering scores, and quantify the sensitivity gain when we apply this simple correction to the overlaps.

In Section \ref{sec:HoleFilling} we explain how to null the effect of identified abrupt noise transients on matched filtering scores of templates for long GW signals. 
We achieve this by replacing (or `inpainting') the bad segments of strain data with values (that we solve for), such that the inverse-PSD-filtered (i.e., twice-whitened, or `blued') data is zero at all bad times.
This guarantees that the offending data has zero influence on any computation of the likelihood, and preserves optimal sensitivity for any real GW event in the surrounding good data.

The methods described in this paper have been implemented in the search pipeline described in Ref.~\cite{pipelinepaper}, together with other improvements in candidate ranking \cite{O2BBHs}, signal consistency checks \cite{vetopaper} and template bank subdivision \cite{templatebankpaper}.

%%%%%%%%%%%%%%%%%%%%%%%%%%%%%%%%%%%%%%%%%%%%%%%%%%%%%%%%%%%%%%%%%%%%%%%
\section{Variations in the noise PSD}
\label{sec:WrongPSD}
%%%%%%%%%%%%%%%%%%%%%%%%%%%%%%%%%%%%%%%%%%%%%%%%%%%%%%%%%%%%%%%%%%%%%%%

In this section we demonstrate that the noise PSD of the LIGO data exhibits slow drifts over timescales of more than a few seconds, and discuss the loss in sensitivity due to this drift. 
We present a simple way to mitigate this problem and quantify the gain in search volume from the proposed correction.

%%%%%%%%%%%%%%%%%%%%%%%%%%%%%%%%%%%%%%%%%%%%%%%%%%%%%%%%%%%%%%%%%%%%%%%
\subsection{PSD estimation and matched filtering}
\label{sec:PSDEstimation}
%%%%%%%%%%%%%%%%%%%%%%%%%%%%%%%%%%%%%%%%%%%%%%%%%%%%%%%%%%%%%%%%%%%%%%%

We begin by defining and briefly reviewing the essential features of matched filtering for signals within data with stationary Gaussian random noise. The statistical properties of the noise are completely described by its autocorrelation function, $C_{\rm n}(\tau)$, defined by
\begin{align}
  \langle d(t)\,d(t + \tau) \rangle & = C_{\rm n}(\tau). \label{eq:autocorr}
\end{align}
The stationary nature manifests in the fact that the function $C_{\rm n}$ is only a function of the lag $\tau$. Additionally, $C_{\rm n}(\tau)$ decays to zero for large values of the lag. The two-sided Power Spectral Density (PSD) is defined as the continuous Fourier transform of the autocorrelation function, i.e.,
\begin{align}
  S_{{\rm n}, 2}(f)  & = \intr d\tau \, C_{\rm n}(\tau)\, e^{-2 \pi i f \tau}. \label{eq:psddef}
\end{align}
The data is real-valued, and hence $C_{\rm n}(\tau)$ and $S_{{\rm n}, 2}(f)$ are real-valued and even functions of their arguments. It is conventional to define the one-sided PSD as $S_{\rm n}(f) = S_{{\rm n}, 2} (f) + S_{{\rm n}, 2} (-f) = 2\, S_{{\rm n}, 2} (f)$.

We are interested in searching for a signal, say $h(t)$, within the data. 
In our use-case, we can take $h(t)$ to have compact support when restricted to the sensitive band of the detectors. 
We work with data that is sampled at an interval $\Delta t$, chosen such that the Nyquist frequency $f_{\rm s}/2 = 1/(2\,\Delta t)$ is high enough to encompass the sensitive band. 

Let us consider a segment of data, of length $N$, that is long enough to contain the putative signal. 
In the absence of the signal, the Discrete Fourier Transform (DFT) of the segment satisfies\footnote{Our convention for the DFT is 
\begin{equation*}
  \tilde{d} \lt f_m \rt = \sum_{n=0}^N\,d(n \Delta t)\,e^{-2 \pi i n \Delta t f_m},
\end{equation*} where 
\begin{equation*}
  f_m = \frac{m}{N\Delta t}, \quad -N/2 + 1 \leq m \leq N/2, 
\end{equation*} with $N$ being even. \label{fn:dft}}
%\begin{multline}
%  \langle \tilde{d}(f_n) \tilde{d}^\ast(f_{n^\pr}) \rangle = \frac12 e^{\pi i (f_n - f_{n^\pr}) (N - 1) \Delta t} \int {\rm d}f S(f) \times \\
%  \lt \frac{\sin{[\pi (f + f_{n^\pr}) N \Delta t]}}{\sin{[\pi (f + f_{n^\pr}) \Delta t]}} \rt \lt \frac{\sin{[\pi (f + f_{n}) N \Delta t]}}{\sin{[\pi (f + f_{n}) \Delta t]}} \rt,
%\end{multline}
\begin{align}
  & \llangle \tilde{d} \lt f_m \rt \left[ \tilde{d} \lt f_{m^\pr} \rt  \right]^\ast \rrangle 
  = \frac12\, e^{i\,\pi\, \lt f_{m^\pr} - f_m \rt\, \Delta t} \notag \\
  & \times \intr {\rm d} f\, S_{{\rm n},2}(f)\, W_N \lt f; f_m \rt W_N \lt f; f_{m^\pr} \rt, \label{eq:psddefdiscrete}
\end{align}
where the window function $W_N \lt f; f_m \rt$ is defined in Eq.~\eqref{eq:window} (see Appendix~\ref{ap:varyingnoisecov} for a derivation). 
When viewed as a function of frequency $f$, $W_N \lt f; f_m \rt$ exhibits a series of peaks with height $N$ and width $\sim 1/(N\Delta t)$, separated by the sampling frequency. 
If we assume that the data was properly bandpassed before sampling (to prevent aliasing), we can restrict to the frequency interval between $-f_{\rm s}/2$ and $f_{\rm s}/2$. 
If the PSD behaves smoothly on frequency scales of $\sim 1/(N\Delta t)$, $W_N \lt f; f_m \rt$ behaves like a delta-function selecting the frequency $f = -f_m$ in the integrand, and we have
\begin{align}
  \llangle \tilde{d} \lt f_m \rt \left[ \tilde{d} \lt f_{m^\pr} \rt \right]^\ast \rrangle \approx \frac{N}{2\,\Delta t}\, S_{\rm n}\lt f_m \rt\,\delta_{m, m^\pr}. \label{eq:psdmeanvar}
\end{align}
It is convenient to define a whitened data stream, $d_{\rm w}$, according to:
\begin{align}
  \wt{d_{\rm w}} \lt f_m \rt & = \left[ \frac{2\,\Delta t}{S_{\rm n} \lt f_m \rt} \right]^{1/2} \tilde{d} \lt f_m \rt, \label{eq:whitedata}
\end{align}
that satisfies
\begin{align}
  \llangle \wt{d_{\rm w}} \lt f_m \rt \left[ \wt{d_{\rm w}} \lt f_{m^\pr} \rt \right]^\ast \rrangle & = N\,\delta_{m, m^\pr}. \label{eq:psdmeanvar_white}
\end{align}
The whitened data is the result of convolving the raw data stream with a whitening filter. We define the convolution of the signal and the whitening filter as the `whitened signal', $h_{\rm w}$:
\begin{align}
  \wt{h_{\rm w}} \lt f_m \rt & = \left[ \frac{2 \Delta t}{S_{\rm n} \lt f_m \rt} \right]^{1/2} \tilde{h} \lt f_m \rt. \, \label{eq:whitesignal}
\end{align}
Under the assumption of stationary Gaussian noise, the matched filtering score, defined as
\begin{align}
    z & = \left[ d_{\rm w} \conv \rev{h_{\rm w}} \right](0) = \frac{1}{N} \sum_m \left[ \wt{h_{\rm w}} \lt f_m \rt \right]^\ast \wt{d_{\rm w}} \lt f_m \rt \label{eq:matchedFilter}
\end{align}
is the optimal detection statistic for the signal $h(t)$ in the data (in the first equation, the symbol $\conv$ represents convolution and the arrow signifies time-reversal). If the data $d$ and signal $h$ are real-valued, so is the score $z$. In the absence of a signal, the variance of the score is
\begin{align}
  \langle z^2 \rangle & = \left[ h_{\rm w} \conv \rev{h_{\rm w}} \right](0) = \frac{1}{N} \sum_m \llvert \wt{h_{\rm w}} \lt f_m \rt \rrvert^2, \label{eq:snr2}
\end{align}
which also equals its expectation value in the presence of the signal. Hence the quantity in Eq.~\eqref{eq:snr2} is the squared signal-to-noise ratio $({\rm SNR}^2)$.
We can search for signals with different arrival times by enumerating over shifts in $h(t)$ in Eq.~\eqref{eq:matchedFilter} using an implementation of the Fast Fourier Transform (FFT). 
In practice, we also search for signals with constant phase-offsets, in which case the template $h(t)$ is complex-valued, and the detection statistic is the absolute value of Eq.~\eqref{eq:matchedFilter}. 
For the sake of simplicity, we assume that the template is real; all the results of this paper hold for complex templates as well. 

%Where $d$ is the raw strain data, $S_n(f)$ is the noise power spectral density, and it is understood that whenever the template, data or PSD are evaluated with a frequency their discrete Fourier transform is used.
%Here, the template, like the data is assumed to be real valued. Although it is sometimes useful to consider a complex template, and to use the absolute value of Equation \ref{eq:matchedFilter} as a detection statistic, we will assume, for the sake of simplicity of explanations, that the template is real.
%All the results of the paper hold also for complex templates. 

In deriving Equation \eqref{eq:matchedFilter} and in proving its optimality, it is assumed that the noise PSD, $S_{\rm n}(f)$, is known, while in practice, we have to measure it from the data itself.
A standard way to do so is the Welch method, which divides the data into many (ideally overlapping) segments, applies a window function followed by a DFT to each segment, and calculates $S_{\rm n}(f)$ as the average of the power spectra in all the segments.
%The criteria for deciding on the number of segments and their duration are the following:

The frequency resolution and the precision of the measured PSD are important criteria to decide the number and duration of segments to use with the Welch method.
The frequency resolution depends on the length of the segments, through the window functions $W_N \lt f; f_m \rt$ defined in Eq.~\eqref{eq:window}.
The advanced LIGO noise contains several sharp spectral lines, at which the PSD is several orders of magnitude higher than the `floor'.
If the chosen segments are too short, the lines are broadened (through convolution with $W_N \lt f; f_m \rt$) and bleed into surrounding frequency bins: the effect is to reduce the SNR, and move us away from optimality.
% Ideally, all sharp spectral features should be resolved by the chosen segment length.
The stochastic error of the PSD measurement depends on the number of segments that are averaged over. 
As shown in the next section, the signal recovery efficiency in the presence of stochastic PSD errors is roughly $1-0.5 \, N_{\rm seg}^{-1}$, where $N_{\rm seg}$ is the number of segments used when estimating a PSD using the Welch method. 
%\mz{I don't understand what this refers to.} \barak{Is it OK now?} \teja{Where is this shown?}
Together, these parameters define a minimal duration over which the PSD needs to be estimated to achieve a target recovery efficiency.
In order to bound the sensitivity loss to less than a few percent, for LIGO data, the duration of the PSD measurement needs to be roughly \SI{e3}{\second}.

%\subsubsection{Evidence for fast changing PSD in LIGO O1 data}

The discussion so far has been theoretical; we would like to test our assumptions, and whether the above procedure achieves the required bounds on the sensitivity. 
The most direct check is to verify the statistics of the matched filtering scores, since they determine our sensitivity. 
In what follows, we normalize the template $h$ so that the variance $\langle z^2 \rangle$ as given by Eq.~\eqref{eq:snr2} equals unity.

\begin{figure}
    \centering
    \includegraphics[width=\linewidth]{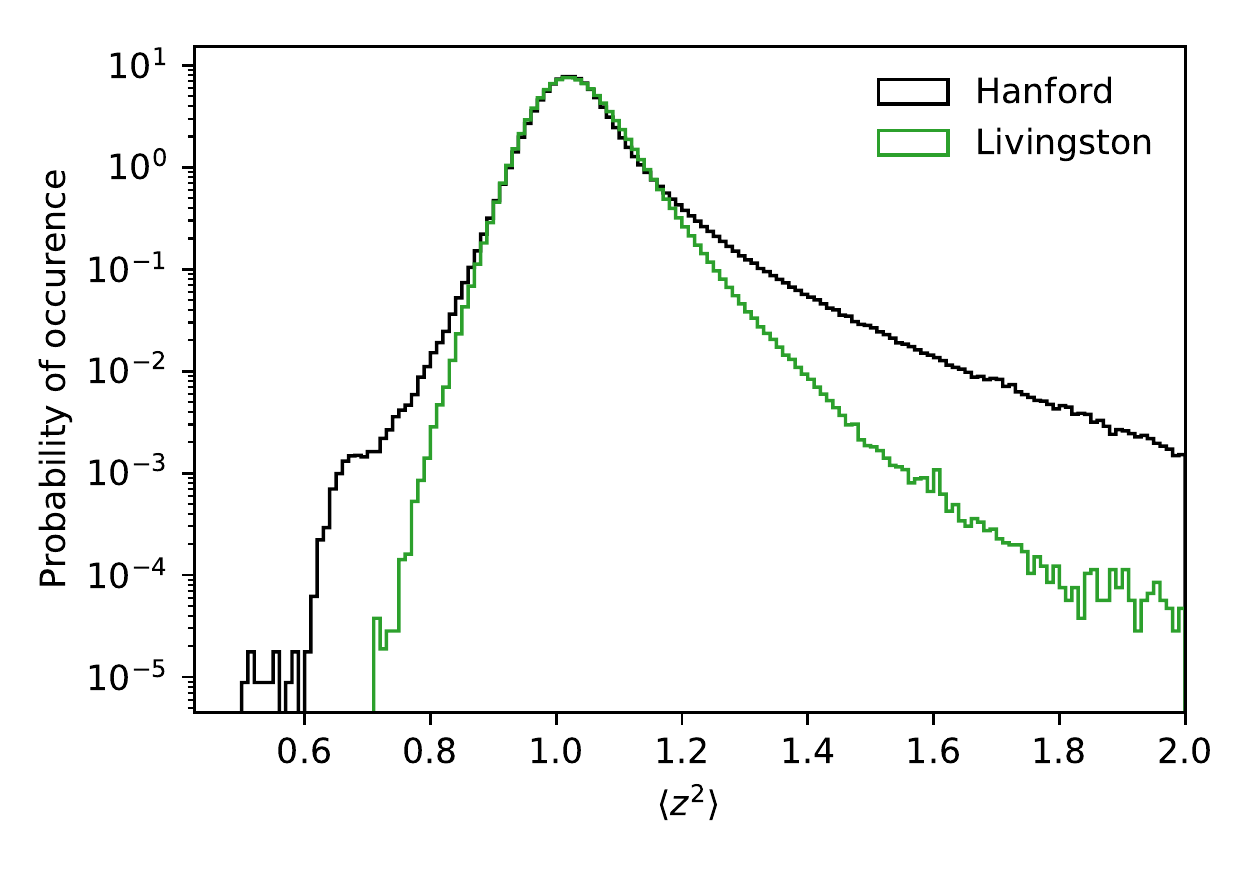}
    \caption{Histogram of variance values of overlaps computed on the entirety of O2 data. Measurement error on these values is $2\%$. Measured deviations from unity are much bigger (about 10\%) and therefore produce a big difference in significance determination if unaccounted for.\label{fig:hist_psd_drift}}
\end{figure}

We consider the matched filtering scores for a given template $h(t)$ as a time-series, i.e., 
\begin{align}
  z(t = n \Delta t) & = \left[ d_{\rm w} \conv \rev{h_{\rm w}} \right](t) \notag \\
  & = \frac{1}{N} \sum_m \left[ \wt{h_{\rm w}} \lt f_m \rt \right]^\ast \wt{d_{\rm w}} \lt f_m \rt e^{2 \pi i f_m n \Delta t}. \label{eq:matchedFiltertimeseries}
\end{align}
The simplest check is whether the actual variance of the scores (as estimated from the time-series) is consistent with unity.
%Apart from the spectrum of the variations in the PSD $\epsilon$, it is also instructive to consider their point-distribution. 
Figure \ref{fig:hist_psd_drift} shows the histogram of the estimated variance of the scores $z$ for a single template over the O2 data. 
We estimated the variance by averaging the power $z^2$ within rolling windows of length $\sim \SI{15}{\second}$, which should achieve $2\%$ stochastic error on the variance.
Instead, we see that the standard deviation of the variance distribution is approximately $8-9 \%$.
In the next section, we investigate the associated loss in sensitivity in more detail.
We now investigate the cause of the phenomenon shown in Fig.~\ref{fig:hist_psd_drift}.

%the point distribution of the scores, $z$: the prediction is that they should be distributed according to the standard normal distribution. 

The derivation of the variance in Eq.~\eqref{eq:snr2} depends on the noise being described by the PSD $S_{\rm n}(f)$, i.e., the whitened data satisfying Eq.~\eqref{eq:psdmeanvar_white}.
The failure in Fig.~\ref{fig:hist_psd_drift} suggests that the PSD we estimated did not whiten the data perfectly.
It is well known that the behavior of the detector varies with time (a drastic example of this is the scaling of the noise curve with the varying level of human activity in the vicinity of the detector).
Due to these phenomena, the noise characteristics can change over timescales that are shorter than the $\mathcal O(10^3)$ seconds we use to measure the PSD.

We can view the variance of this series (or instantaneous power, $z^2$) itself as a time-series, that is described by its own PSD, $S_{z^2}$. In the stationary case (when Eq.~\eqref{eq:psdmeanvar} holds), $S_{z^2}$ equals
\begin{align}
  S_{z^2} \lt f_m \neq 0 \rt & = 4 \Delta t \, \wt{\llvert h_{\rm w} \conv \rev{h_{\rm w}} \rrvert^2} \lt f_m \rt, \label{eq:sz2}
\end{align}
where $h_{\rm w} \conv \rev{h_{\rm w}}$ is the autocorrelation function of the whitened waveform (see Appendix \ref{ap:powerspectrumvar} for a derivation). %Figure \ref{fig:wf_acorr} shows the autocorrelation function for a typical waveform in our \texttt{BBH 3} bank (which contains templates for massive black-hole mergers); 
The autocorrelation function of a typical waveform has a width that is of order few ms, and hence at frequencies smaller than \SI{1}{\hertz} we expect the PSD of the power, $S_{z^2}$, to be flat.

\begin{figure}
    \centering
    \includegraphics[width=\linewidth]{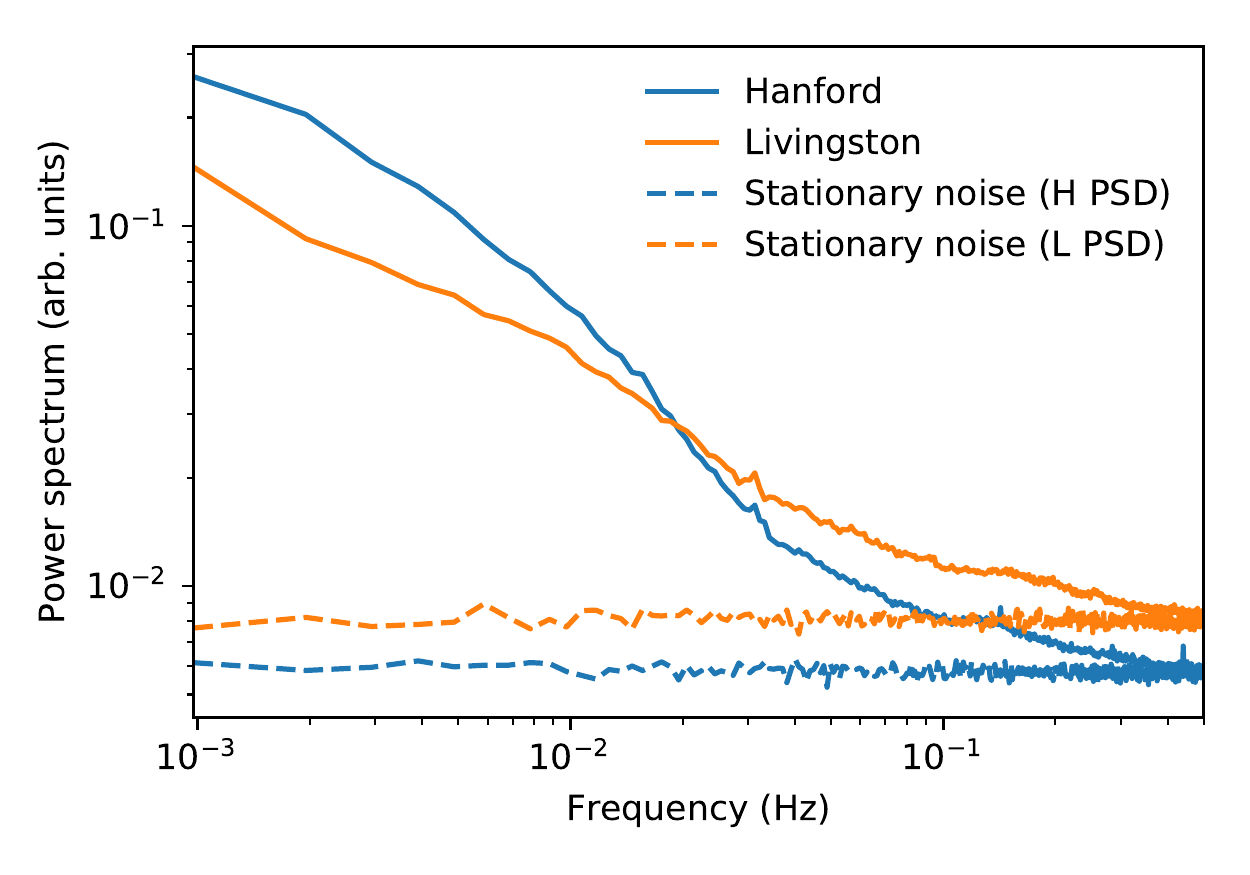}
    \caption{It is necessary to track the drifting PSD on time scales of seconds. Solid lines show the empirically measured power spectrum $S_{z^2}$ of a time-series composed of the measured variance of the overlaps in every second (note that the frequencies are much lower than the frequency content of the waveform itself). For reference, dashed lines show the same on artificially generated stationary Gaussian noise. The variance time-series has a red-noise power spectrum.}
    \label{fig:ps_psd_drift}
\end{figure}

%As a counter to the requirements to resolves the lines and having an accurate PSD measurement, stands the fact that the PSD may change due to many different physical processes in the detector.
Figure \ref{fig:ps_psd_drift} shows the PSD of the variance $S_{z^2}$ of the matched filtering scores, computed using a heavy binary black hole template (we estimate the local variance by convolving the $z^2$ series with a rolling rectangular window of \SI{1}{\second} duration).
The dashed curves are the estimated $S_{z^2}$, for scores measured on stationary Gaussian noise generated using the fiducial PSDs for the Livingston and Hanford detectors, respectively: they are flat as a function of frequency, in line with the prediction of Eq.~\eqref{eq:sz2}.
The solid curves show the empirically measured variance PSDs, $S_{z^2}$, for L1 and H1 data, averaged over the entirety of O2.
We omitted from the average any region flagged as invalid either by the LIGO and Virgo Collaboration, or by our pipeline (see \cite{pipelinepaper}), keeping only the contiguous segments.
In the rest of this section, we describe how non-stationary noise can lead to such red-noise spectra, and what the measured curve tells us about the departure from the stationary case.

In the non-stationary case, we start by generalizing Eqs.~\eqref{eq:autocorr} and \eqref{eq:psddef}:
\begin{align}
  \langle d(T - \tau/2)\,d(T + \tau/2) \rangle & = C_{\rm n}(\tau) + \delta C_{\rm n}(\tau; T), \, \label{eq:nsautocorr} 
\end{align}
and 
\begin{align}
  S_{{\rm n}, 2}(f; T)  & = S_{{\rm n}, 2}(f)\,\left[ 1 + \epsilon(f; T) \right], \, {\rm where} \label{eq:nspsd0} \\
  \epsilon(f; T) & = \frac{1}{S_{{\rm n}, 2}(f)}\,\intr d\tau \, \delta C_{\rm n}(\tau; T)\,e^{-2 \pi i f \tau}. \label{eq:nspsd1}
\end{align}
In the above equations, $\epsilon(f; T)$ is the fractional change in the noise PSD at frequency $f$, which we take to vary on timescales $T \gg \tau$ (the frequency $f$ is conjugate to the short timescale $\tau$). 

If we whiten the data using the PSD estimated assuming stationary noise (i.e., $S_{\rm n}(f)$) over a duration that is longer than the timescales over which the PSD varies, the equivalent of Eq.~\eqref{eq:psdmeanvar_white} is:
\begin{align}
  \!\!\!\! & \llangle \wt{d_{\rm w}} \lt f_m \rt \left[ \wt{d_{\rm w}} \lt f_{m^\pr} \rt \right]^\ast \rrangle & \approx N\,\delta_{m, m^\pr} + \tilde{\epsilon} \lt \bar{f}; \Delta f \rt, \label{eq:nspsdmeanvar}
\end{align}
where $\bar{f} = \lt f_m + f_{m^\pr} \rt/2$, and $\Delta f = f_m - f_{m^\pr}$ (see Appendix \ref{ap:varyingnoisecov} for a derivation). In the second term, $\tilde{\epsilon} \lt \bar{f}; \Delta f \rt$ is the DFT, evaluated at $\Delta f$, of $\epsilon \lt \bar{f}, T = n \Delta t \rt$ sampled at a rate of $f_{\rm s} = 1/(\Delta t)$. Equation \eqref{eq:nspsdmeanvar} was derived under the assumption that the noise PSD, $S_{\rm n}(f)$, is smooth on frequency scales of $\Delta f$, and as such is inaccurate in the immediate vicinity of spectral lines in $S_{\rm n}(f)$.

The non-stationary term, $\epsilon$, correlates Fourier modes of the data, $d \lt f_m \rt$, with different frequencies. Intuitively, if we analyze segments that are shorter than the slow timescale, $T$, and whiten them using the instantaneous PSD, correlations between Fourier modes are diagonal in terms of frequency $f_m$ (as in Eq.~\eqref{eq:psdmeanvar_white}). Over longer timescales, the frequencies of the (slow) PSD variations beat against the frequencies of the (fast) Fourier modes and lead to the second term in Eq.~\eqref{eq:nspsdmeanvar}.

We are interested in the effect of the non-stationary part of the noise PSD, $\epsilon$, on the PSD of the power in the matched filtering overlaps, $S_{z^2}$.
For this, it is useful to view the values of $\epsilon \lt f_m, T \rt$ themselves as being drawn from a set of random time-series (one for each `fast' frequency $f_m$).
The simplest model for this series is that all the $\epsilon \lt f_m, T \rt$ vary in step with each other, and with the same amplitude (it is a straightforward generalization to model more complicated behavior).
In this case, there is a single PSD that describes the variations:
\begin{align}
  \llangle \tilde{\epsilon} \lt f_m, f_a \rt \left[ \tilde{\epsilon} \lt f_{m^\prime}, f_b \rt \right]^\ast \rrangle & = \frac{N}{2\,\Delta t}\,S_{\epsilon} \lt f_a \rt\,\delta_{a, b}. \label{eq:epsilonpsd}
\end{align}
We can obtain a simple form for the corrected version of Eq.~\eqref{eq:sz2} under additional approximations: (a) variations in the noise PSD (described by $S_\epsilon$) have support at much lower frequencies than the whitened signal $\wt{h_{\rm w}}$ does, and (b) the time-domain whitened waveform is much shorter than the timescales over which $\epsilon(f; T)$ varies. In this case, we have
\begin{align}
  \!\!\!\! & S_{z^2} \lt f_m \neq 0 \rt \notag \\
  & \approx 2 \Delta t \, \wt{\llvert h_{\rm w} \conv \rev{h_{\rm w}} \rrvert^2} \lt f_m \rt \left[ 2 + \intr {\rm d}f \, S_{\epsilon} \lt f \rt \right] + \notag \\
  & ~~~\llvert \wt{h_{\rm w}^2} \lt f_m \rt \rrvert^2 \, S_\epsilon \lt f_m \rt. \label{eq:ns_sz2}
\end{align}
Appendix \ref{ap:powerspectrumvar} presents a detailed  derivation of this equation. The first term in Eq.~\eqref{eq:ns_sz2} is flat with frequency at low frequencies (similarly as in Eq.~\eqref{eq:sz2}). The second term is proportional to the power spectrum, $S_\epsilon$, of the non-stationary part of the noise-PSD, $\epsilon$. 
The behavior of the solid curves in Fig.~\ref{fig:ps_psd_drift} suggests that that the PSD, $S_{\rm n}(f; T)$, itself varies on timescales $T$ larger than a few seconds, and that these variations have a red spectrum.

%As an indication to the changes in the PSD, we measure the local variance of the overlaps $z$ computed with a binary black hole merger template. (The results are identical between different templates with the same $|h(f)|$ and hence the particular template does not matter). We compute the local variance by averaging $z^2$ over a window of 1 second in duration. We then treat this time-series as a stochastic process and compute its power spectrum. In Figure \ref{fig:ps_psd_drift} we show the power spectrum of the variance measurement. The power spectrum indicates that substantial PSD changes are occurring from times that are as short as a few seconds.

%%%%%%%%%%%%%%%%%%%%%%%%%%%%%%%%%%%%%%%%%%%%%%%%%%%%%%%%%%%%%%%%%%%%%%%
\subsection{The loss of sensitivity due to a wrong PSD}
\label{sec:PSDEsensitivityloss}
%%%%%%%%%%%%%%%%%%%%%%%%%%%%%%%%%%%%%%%%%%%%%%%%%%%%%%%%%%%%%%%%%%%%%%%

Suppose that instead of the true PSD, $S_c(f)$, we use a wrong one due to PSD misestimation: 
\begin{align}
    S_w(f) = S_c(f)\,(1 + \epsilon(f))\,.
\end{align} 
In this section, we compute the bias to the recovered SNR of a putative event due to the wrong PSD. 
We will show that:
\begin{enumerate}
    \item The first-order effect of using a wrong PSD in computing Eq.~\eqref{eq:matchedFilter} is mis－estimation of its standard deviation.
    \item We show that if the standard deviation of the overlaps is corrected, then the SNR loss will be of order $\epsilon^2(f)$.
\end{enumerate}
%\begin{equation}
%     S_c(f)  = S_w(f) + \delta S(f)
%\end{equation}
Let us consider the following statistical model for the strain data:
\begin{equation}
    d(f) =\alpha\, h(f) + n(f),
\end{equation}
where $h(f)$ is a compact binary coalescence template waveform with an amplitude normalization $\alpha$, and $n(f)$ is the noise which obeys the true PSD $S_c(f)$.
For simple notation, we define
\begin{align}
    I(f)\equiv \left| h(f)\right|^2/S_c(f),   
\end{align}
which is normalized according to
\begin{align}\label{eq:infoNorm}
    \sum_f{I(f)} = 1.
\end{align}
We now compute the overlap using the wrong PSD:
\begin{align}
    z_w = \sum_{f}{\frac{h^*(f)\,d(f)}{S_w(f)}}\,.
\end{align}
In the absence of a signal $\alpha=0$, this overlap has a variance $\lambda_w$:
\begin{equation}
    \begin{split}
        \lambda_w &\equiv \langle|z_w|^2\rangle \\ %- \left|\langle z_w \rangle_c\right|^2 \\
        &= \sum_{f}{\frac{h^*(f)\,h(f)}{S_w^2(f)}}\,S_c(f)\\
        &= \sum_f{\frac{I(f)}{(1+\epsilon(f))^2}}.
    \end{split}
\end{equation}
We note that this is the true variance of $z_w$, which is not easily computable as it depends on the unknown $S_c$. We will however measure it directly from data. 

The naive way to calculate the variance involves using the wrong PSD $S_w$:
%If we calculate (as was the standard practice until now) the variance using $S_w$ alone, we get:
\begin{equation}
\begin{split}
    \lambda_{w\,{\rm computed}} &\equiv \sum_{f}{\frac{h^*(f)\,h(f)}{S_w(f)}} \\
    &=\sum_f{\frac{I(f)}{1+\epsilon(f)}}\,.
\end{split}
\end{equation}
In the presence of a signal, the expected response is given by:
\begin{align}
    \alpha_w \equiv \langle z_w \rangle_h = \alpha \sum_f{\frac{|h(f)|^2}{S_w(f)}} =  \alpha \sum_f{\frac{I(f)}{1+\epsilon(f)}}
\end{align}
Thus, if we use $z_w$ as our statistic, the significance squared is 
\begin{equation}
  \rho_w^2 = \frac{\langle z_w \rangle^2_h}{\langle|z_w|^2\rangle} = \frac{\alpha_w^2}{\lambda_w}\,.  
\end{equation}
If we were to take for granted that the variance of $z_w$ was $\lambda_{w\,{\rm computed}}$, we would have estimated:
\begin{equation}
  \rho_{w,{\rm computed}}^2 = \frac{\alpha_w^2}{\lambda_{w\,{\rm computed}}}\,.  
\end{equation}

If the true PSD is known, the optimal overlap is given by:
\begin{align}
    z_c = \sum_{f}{\frac{h^*(f)\,d(f)}{S_c(f)}}\,
\end{align}
which maximizes the SNR. In the presence of a signal:
\begin{align}
    \alpha_c \equiv \langle z_c\rangle_h = \alpha \sum_f{\frac{|h(f)|^2}{S_c(f)}} = \alpha \sum_f{I(f)} = \alpha.
\end{align}
Absent the signal, $z_c$ has a variance
\begin{equation}
    \lambda_c \equiv \langle|z_c|^2\rangle = \sum_{f}{\frac{|h(f)|^2}{S_c^2(f)}}\,S_c(f) = \sum_f{I(f)} = 1.
\end{equation}
The optimal $\rho_c^2$ is then:
\begin{equation}
  \rho_c^2 = \frac{\langle z_c \rangle^2_h}{\langle|z_c|^2\rangle} = \frac{\alpha_c^2}{\lambda_c} = \alpha^2\,.
\end{equation}
Therefore, the relative information loss between using the true PSD $S_c(f)$ and the wrong PSD $S_w(f)$ is given by:
\begin{align}
\label{eq:SNR2c_to_SNR2w}
    \frac{\rho_c^2}{\rho_w^2}
    = \frac{\alpha_c^2\,\lambda_w}{\alpha_w^2\,\lambda_c}
    = \frac{\sum_f{I(f)/(1+\epsilon(f))^2}}
           {\big[\sum_f{I(f)/(1+\epsilon(f))}\big]^2}.
\end{align}
If $\epsilon(f)$ was frequency independent, then the ratio would be one, and $\rho_w^2$ would be as optimal as $\rho_c^2$. More generally, $\rho_c^2 > \rho_w^2$ for any frequency dependent $\epsilon(f)$, so there is always a sensitivity loss. 

We now Taylor expand \eqref{eq:SNR2c_to_SNR2w}) and keep terms up to second order in $\epsilon(f)$:
\begin{align}
    \frac{\rho_c^2}{\rho_w^2}
    &= \frac{1-2\,\sum_f{I(f)\,\epsilon(f)}+3\,\sum_f{I(f)\,\epsilon^2(f)}}
        {\left[ 1-\sum_f{I(f)\,\epsilon(f)}
         + \sum_f{I(f)\,\epsilon^2(f)} \right]^2} \\ \nonumber
    &= 1 + \sum_f{I(f)\,\epsilon^2(f)}
       - \Big(\sum_f I(f)\,\epsilon(f)\Big)^2 \geq 1.
\end{align}
% Again to simplify our notation let us re-express the change in the PSD as:
% \begin{equation}
%      S_c(f)  = \lambda S_w(f) [1 + \epsilon(f)],
% \end{equation}
% which using the definition of $\lambda$ can be shown to satisfy:
% \begin{equation}
%     \sum_{f}{\frac{h^*(f)h(f)}{S_w(f)}} \epsilon(f) =0. 
% \end{equation}
% We can now ask what would happen if we had used the correct PSD to compute the scores:
% \begin{align}
%     z_c = \sum_{f}{\frac{h^*(f)d(f)}{S_c(f)}}.
% \end{align}
% In this case the $SNR_c^2$ is 
% \begin{equation}
% \begin{array}{rcl}
%         SNR_c^2 &=& \frac{\langle z_c \rangle^2}{\langle|z_c|^2\rangle} \\
%          &=& \alpha^2 \sum_{f}{\frac{h^*(f)h(f)}{S_w(f)}} \\
%          &=& \alpha^2 \sum_{f}{\frac{h^*(f)h(f)}{\lambda S_w(f) [1 + \epsilon(f)]}} \\
%          &\approx& \frac{\alpha^2}{\lambda} \sum_{f}{\frac{h^*(f)h(f)}{S_w(f)}} 
%          [1 - \epsilon(f)+ \epsilon(f)^2 + \cdots] \\
%          &\approx& \frac{\alpha^2}{\lambda} (1 + \sum_{f}{\frac{h^*(f)h(f)}{S_w(f)}} \epsilon(f)^2 + \cdots)
% \end{array}
% \end{equation}
Thus, the leading correction to $\rho^2$ is quadratic with $\epsilon(f)$.
This is understandable from the fact that the overlap $z_c$ is constructed to maximize the SNR, and thus loss of sensitivity due to an error in PSD estimation should be a quadratic function.

The relative error in PSD estimation in each frequency band of $S_c(f)$ is $0.7\,N_{\rm seg}^{-1/2}$, where $N_{\rm seg}$ is the number of segments used for measuring the PSD using the Welch method (the coefficient $0.7$ is determined empirically). Therefore, as a first corollary, the fractional sensitivity loss due to the statistical error in PSD estimation is $0.5\,N_{\rm seg}^{-1}$.

A key point is that computing $\rho_w^2$ requires $\lambda_w$, which cannot be computed without the knowledge about $S_c$. Nevertheless, $\lambda_w$ can be directly measured from the data.
We now compute the fractional error in the standard deviation estimation:
\begin{equation}
    \begin{split}
        \frac{\lambda_w}{\lambda_{w\,{\rm computed}}}
        &= \frac{\sum_f{I(f)/(1+\epsilon(f))^2}}{\sum_f{I(f)/(1+\epsilon(f))}} \\
        &= 1-\sum_f{I(f)\,\epsilon(f)} + \mathcal{O}\left[\epsilon^2(f)\right].
    \end{split}
\end{equation}
At the leading order, the error scales linearly with $\epsilon(f)$. Thus if one uses $\lambda_{w, {\rm computed}}$ to compute the SNR of an event one makes a linear mistake. We can think of $\sum_f[{{I(f)}/{(1+\epsilon(f)})}]$ as a random variable which adds $\mathcal{O}[\epsilon(f)]$ noise to the estimate of the trigger SNRs. As we will describe in the next sections, this creates a tail in the distribution of triggers which results in substantial sensitivity loss. In the next sections we will discuss how to measure $\lambda_w$. As a remedy to PSD misestimation, this mitigates the sensitivity loss from $\mathcal{O}[\epsilon(f)] \sim 10 \%$ to order  $\mathcal{O}[\epsilon^2(f)]\sim 1 \%$.

%%%%%%%%%%%%%%%%%%%%%%%%%%%%%%%%%%%%%%%%%%%%%%%%%%%%%%%%%%%%%%%%%%%%%%%
\subsection{PSD drift correction}
%%%%%%%%%%%%%%%%%%%%%%%%%%%%%%%%%%%%%%%%%%%%%%%%%%%%%%%%%%%%%%%%%%%%%%%

We propose a practical method to mitigate the sensitivity loss due to fast variations in the PSD (to second order in the amplitude of the drift) by tracking the time-dependent variance of the calculated overlaps at few percent accuracy. 
Since this is a single number to be estimated from the data (contrary to a full PSD which consists of thousands of numbers), we can estimate it using very short segments of data. For a 2\% relative error in the variance, we will need roughly \SI{10}{\second} of data. In Appendix \ref{ap:ErrorComputation} we estimate the precision with which this variance can be determined depending on the length of the data segments used.
The variance around any given time $t$ is empirically computed in the time domain via
\begin{align}
\label{eq:timeDomainPSDDrift}
    \lambda_w(t) \equiv \langle|z_w(t)|^2\rangle &\approx \frac{1}{N_{\rm a}}\sum_{t'=t - N_{\rm a}\Delta t/2}^{t + N_{\rm a}\Delta t/2}{|z_w(t')|^2} \\ \nonumber &\propto \sum_{t'=t - N_{\rm a}\Delta t/2}^{t + N_{\rm a}\Delta t/2}{|h_{\rm w}(t')\conv d_{\rm w}(t')|^2} \\ \nonumber &= \sum_f{\bigg|\frac{h^*(f)\,d(f)}{S_w(f)}\bigg|^2} \\ \nonumber &=\sum_f{\frac{|h^*(f)|^2|d(f)|^2}{S_w^2(f)}},
\end{align}
where $N_{\rm a}$ is the number of scores used in the average, and $h_{\rm w}$ and $d_{\rm w}$ are the template and data whitened with the wrong PSD $S_w$, respectively.
The third equality is due to Parseval's theorem. The normalization factor in the above formula depends on the choice of $N_{\rm a}$ and can be precisely worked out.

The approximate equality in \eqref{eq:timeDomainPSDDrift} is due to the finite number of samples used, which results in an uncertainty proportional to $N_{\rm a}^{-1/2}$.
Exact equality does not hold also because the PSD may vary within the averaging timescale $N_{\rm a} \Delta t$. We invite the readers to read Appendix \ref{ap:ErrorComputation} for a full derivation.

It is noteworthy that $\lambda_w(t)$ is independent of the phase of $h(f)$ as long as the waveform has its support within $N_{\rm a} \Delta t$. More generally, a correction that is computed for a particular waveform $h_0(f) = A_0(f)\,e^{i\,\psi_0(f)}$ is valid for any other waveform $h(f)=A(f)\,e^{i\,\psi(f)}$ who shares the same amplitude profile $A(f)=A_0(f)$ and whose time domain support, after dechirping,
\begin{align}
    h_{\rm dechirped}(f) = h(f)\,e^{-i\,\psi_0(f)}
\end{align}
is shorter than $N_{\rm a} \Delta t$.

In choosing the timescale on which the moving average is computed for $\lambda_w(t_0)$, there are two mutually competing factors. One is that we prefer the timescale to be as short as possible in order to capture the temporal variation in the PSD as much as possible. Another is that a smaller timescale for averaging leads to a larger sampling uncertainty in the measurement of $\lambda_w(t)$. We adopt a compromise in which the timescale is chosen to have 2\% sampling uncertainty on $\lambda_w(t)$. We estimate that for this value, the error due to not capturing faster variability is comparable, albeit varies slightly with time and from one interferometer to another.
Another important issue is to ensure that the correction is not influenced by loud and abrupt noise transients or real GW events. We achieve this by computing Eq.~\eqref{eq:timeDomainPSDDrift} using a safe mean, namely we only average over overlaps that do not exceed $4.38 \sigma$.

\begin{figure}
    \centering
    \includegraphics[width=\linewidth]{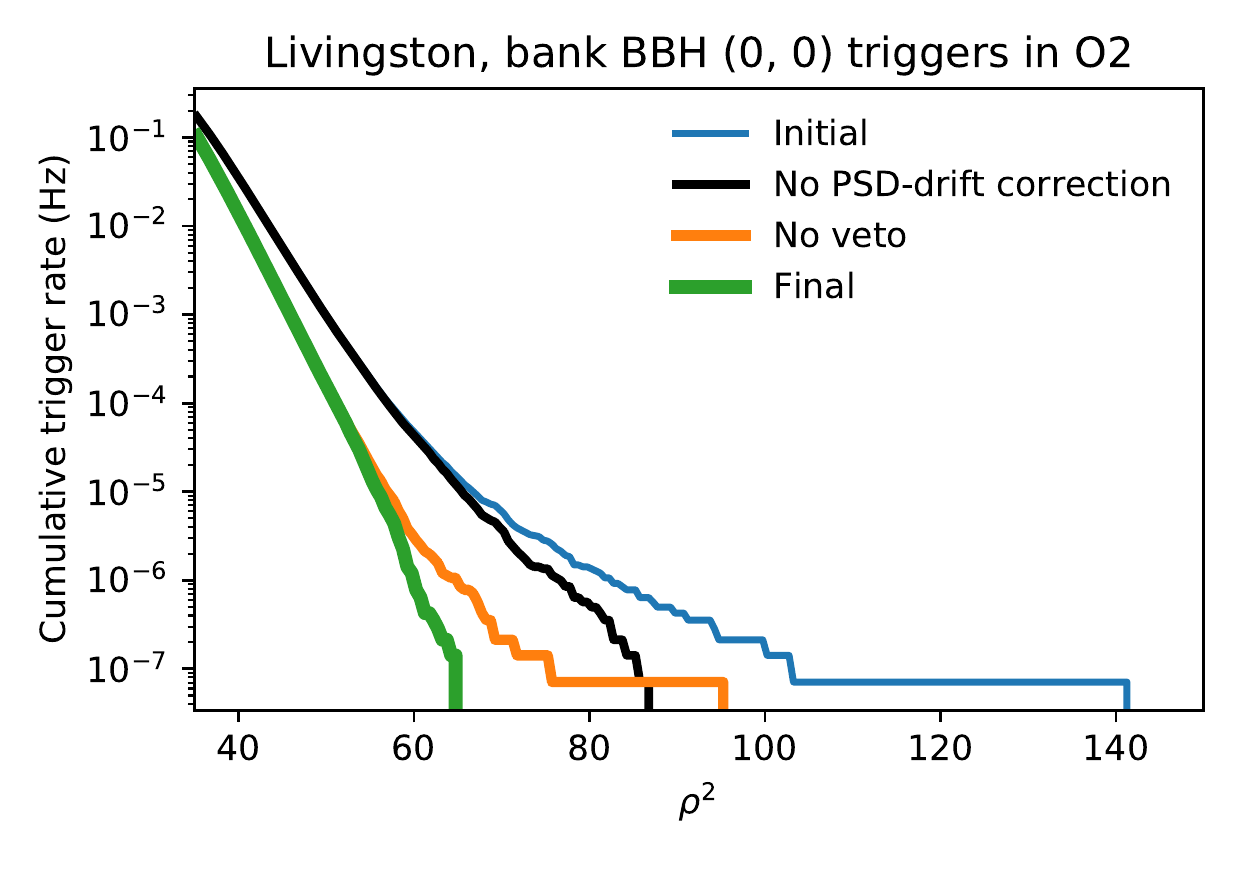}
    \includegraphics[width=\linewidth]{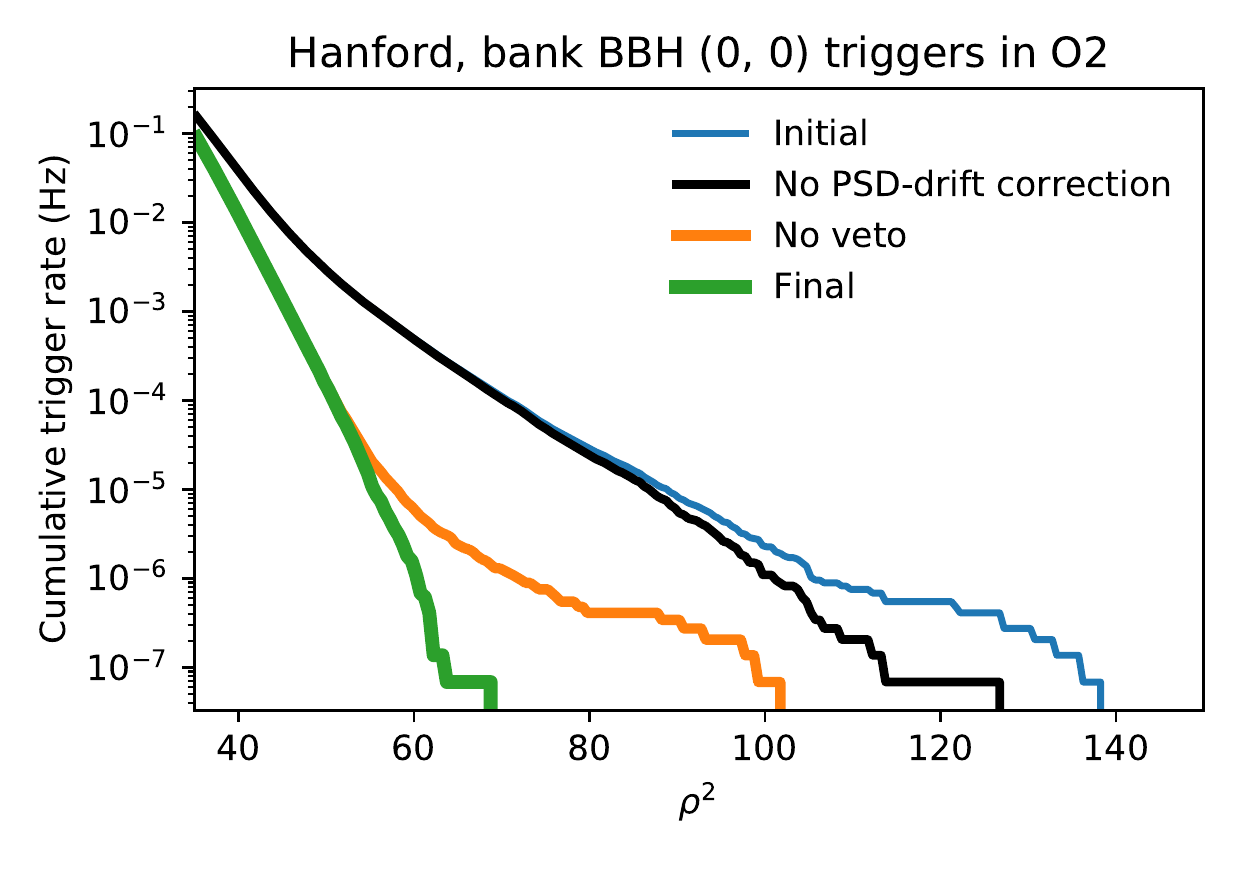}
    \caption{Cumulative trigger rates in bank \texttt{BBH (0,0)} (binary black holes whose chirp mass falls in the range $3$--$5\,M_\odot$~\cite{templatebankpaper}) using the entirety of the O2 bulk data. The initial trigger distribution (blue) is the distribution of the maximum overlap obtained every second, computed after removal of bad data, as described in Ref.~\cite{pipelinepaper}.
    In green is shown the final cumulative trigger distribution, after correcting for PSD drift and applying signal consistency vetoes. In black we show the cumulative trigger distribution when PSD drift is unaccounted for. For reference we show in orange the trigger distribution when the signal consistency vetoes are not applied.
    While real events are in general left untouched by the PSD drift correction (see Fig.~\ref{fig:hist_psd_drift}), the background distribution undergoes a dramatic change. This is because if left unaccounted for, the fluctuations in places with variance misestimation dominate the tail of the trigger distribution. This effect becomes more severe at higher trigger significance. %as higher significance is required from a single detector.
    }
    \label{fig:rescaled_trig_b0}
\end{figure}

\begin{figure}
    \centering
    \includegraphics[width=\linewidth]{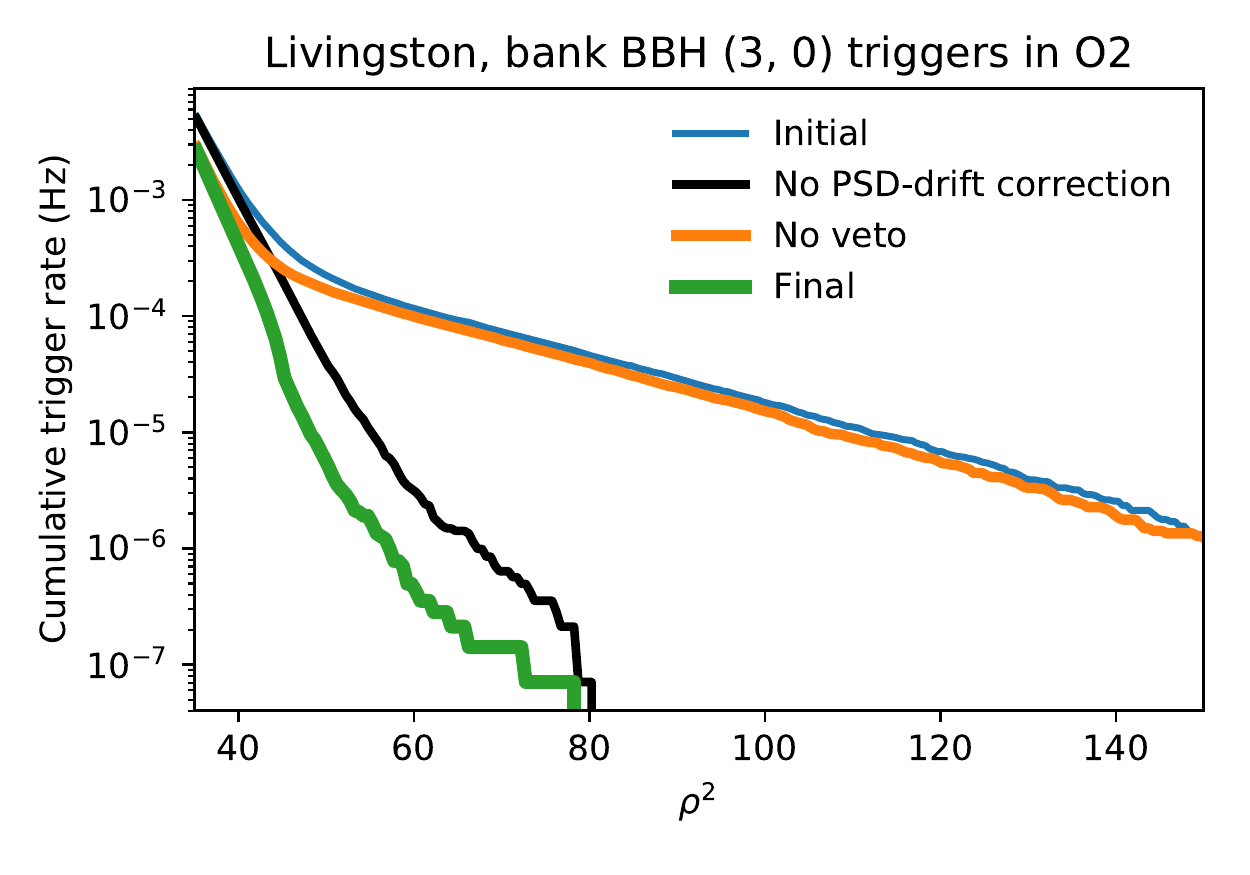}
    \includegraphics[width=\linewidth]{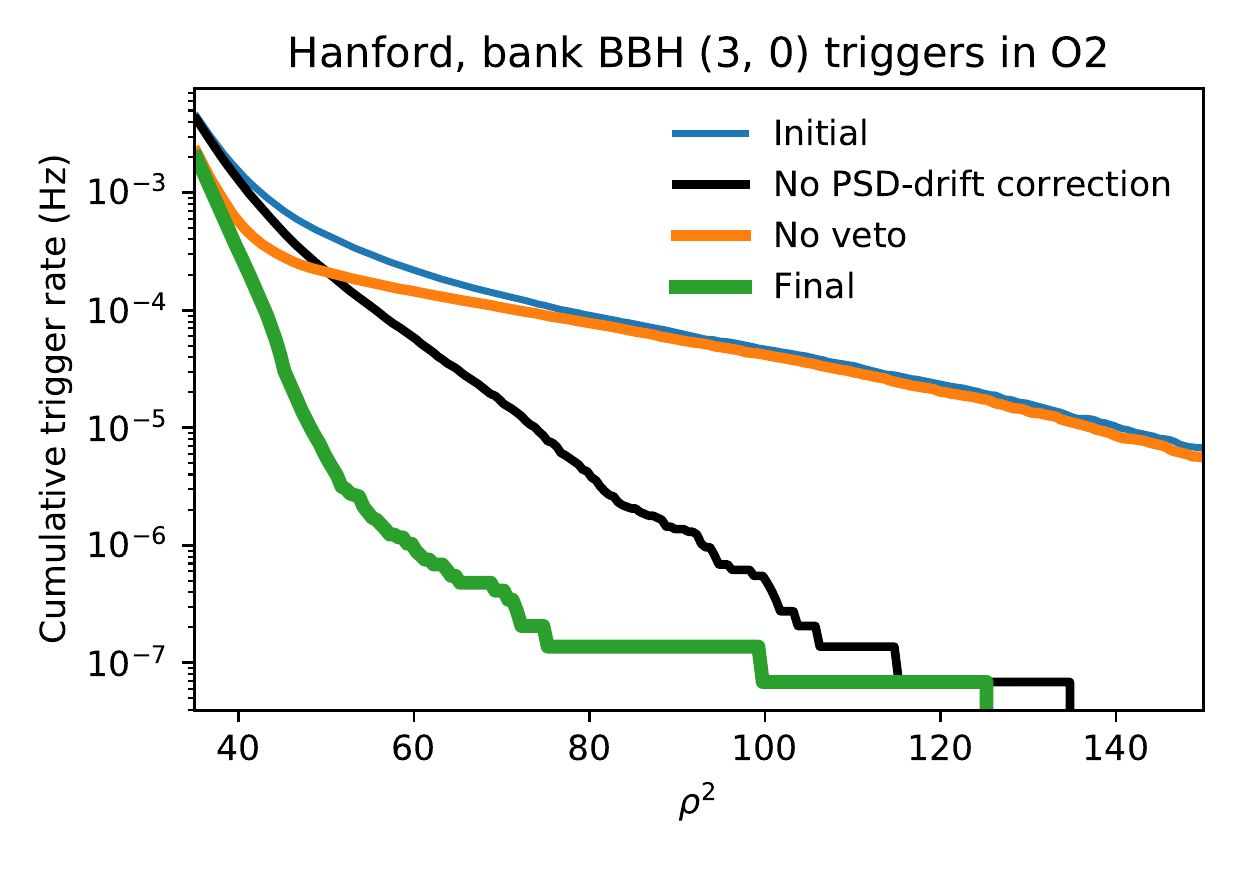}
    \caption{Cumulative trigger rates in bank \texttt{BBH (3,0)} (binary black holes whose chirp mass falls in the range $20$--$40\,M_\odot$ and whose effective spin does not have a very negative value~\cite{templatebankpaper}), using the entirety of the O2 bulk data. This bank contains most of the detected BBH events in O1 and O2. As can be seen, signal consistency checks for triggers are much more important in this domain as the waveforms in general have a very short duration in band. But by pushing back the background distribution, PSD drift correction still mitigates a substantial amount of sensitivity loss.}
    \label{fig:rescaled_trig_b30}
\end{figure}

In Figs.~\ref{fig:rescaled_trig_b0} and \ref{fig:rescaled_trig_b30} we show the dramatic impact on the trigger distribution from applying the PSD drift correction. In the figures we compare the histogram of the PSD-drift-corrected $|z_w|^2/\lambda_w$ statistic and that of the usual $|z_w|^2/\lambda_{w\, \rm computed}$.
The relative importance of this correction and signal consistency vetoes depends on the waveform duration. In Fig.~\ref{fig:rescaled_trig_b0} we show the trigger distribution for our binary black hole bank \texttt{BBH (0,0)} \cite{templatebankpaper}, which contains relatively long waveforms in the chirp-mass range $3-5M_\odot$, here the PSD drift correction is the most important correction.
In Fig.~\ref{fig:rescaled_trig_b30} we use a bank with shorter waveforms, and it can be seen that the PSD drift correction is secondary in importance to signal consistency vetoes. Still, once those have been applied, the PSD drift correction substantially improves the sensitivity.

%%%%%%%%%%%%%%%%%%%%%%%%%%%%%%%%%%%%%%%%%%%%
%%%%%%%%%%%%%%%%%%%%%%%%%%%%%%%%%%%%%%%%%%%%

%%%%%%%%%%%%%%%%%%%%%%%%%%%%%%%%%%%%%%%%%%%%
%%%%%%%%%%%%%%%%%%%%%%%%%%%%%%%%%%%%%%%%%%%%

We quantify the volume improvement by comparing the incoherent double detector trigger distributions from the entirety of O2 in banks \texttt{BBH 0} and \texttt{BBH 3} before and after PSD drift correction. We further restrict to ${\lambda_w}/{\lambda_{w\, \rm computed}} < 1.4$, as large correction values often indicate big disturbances in the spectrogram and the associated seconds are often rejectable by other tests. We also restricted the results to places were both detectors had comparable response $(20 < (z^2/\lambda)_{\rm H, L} < 50)$, where H, L refer to the LIGO detectors at Hanford and Livingston.
We determine a 15\% volume increase in both banks due to correcting the PSD drift effect. This is estimated using
\begin{equation}
    \begin{split}
    \frac{V_{\rm corr}}{V_{\rm uncorr}} 
    &= \left[\ddfrac{\left(\frac{z^2_{\rm H}}{\lambda_{\rm H}} + \frac{z^2_{\rm L}}{\lambda_{\rm L}}\right) \bigg|_{\rm FAR = 1/O2}}
        {\left(\frac{z^2_{\rm H}}{\lambda_{\rm H\,computed}} + \frac{z^2_{\rm L}}{\lambda_{\rm L\,computed}}\right) \bigg|_{\rm FAR = 1/O2}}\right]^{-3/2}
     \\
    &\approx 1.15,
    \end{split}
\end{equation}
where the terms are evaluated at the threshold where the background distribution produces a false alarm rate of 1 trigger per O2 run.
We expect this volume increase to be the same for BNSs as well as for NSBHs. This is a conservative lower bound as many real events would not have comparable detector response, especially in the latest observing runs where the sensitivity greatly differs between detectors. In the case where the detection practically hinges on one detector alone, the volume increase is 50\% for Livingston and 150\% for Hanford, as can be seen in Figs.~\ref{fig:rescaled_trig_b0} and \ref{fig:rescaled_trig_b30}. Since for a substantial amount of time in the second and third observing runs there is great asymmetry between the sensitivity of the detectors, as well as due to natural geometric considerations (and noise fluctuations), the volume contribution from this regime can be substantial.
 
% %\subsection{Potential increase in sensitivity by understanding the source of PSD drift}

% Figure \ref{fig:rescaled_trig_coinc_b3} shows the effect in our coincident triggers in Bank 3. As an example we can consider the candidate we reported in \cite{GW151216} which had a corrected $SNR^2 = 74$. In the corrected distribution the same probability is reached at $SNR^2 = 81$. This corresponds to an increase in horizon volume of $(81/74)^{3/2} \approx 1.15$. If rather than compare the place with the same probability we look at the cumulative distribution and demand those to be the same, the required $SNR^2 = 84.4$ which leads to a volume increase of $(84.4/74)^{3/2} \approx 1.2$. We note that in the case of lower chirp-mass banks, the volumetric improvement due to PSD drift is more dramatic, and is generally increasing with detector asymmetry.
% At the limit of single detector detection, PSD drift improves the detection volume by up to 1.6.

% \begin{figure*}
%     \centering
%     \includegraphics[width=\textwidth]{rescaled_triggers_coinicident.pdf}
%     \caption{There is a conceptual problem with this Figure, the "Before correction" population is under represented. Solution: Limit correction to 1.4, limit y axis to 0.1 (cumulative)}
%     \label{fig:rescaled_trig_coinc_b3}
% \end{figure*}

%%%%%%%%%%%%%%%%%%%%%%%%%%%%%%%%%%%%%%%%%%%%%%%%%%%%%%%%%%%%%%%%%
\section{Hole filling} 
\label{sec:HoleFilling}
%%%%%%%%%%%%%%%%%%%%%%%%%%%%%%%%%%%%%%%%%%%%%%%%%%%%%%%%%%%%%%%%%

%%%%%%%%%%%%%%%%%%%%%%%%%%%%%%%%%%%%%%%%%%%%%%%%%%%%%%%%%%%%%%%%%
\subsection{Signal processing rationale}
%%%%%%%%%%%%%%%%%%%%%%%%%%%%%%%%%%%%%%%%%%%%%%%%%%%%%%%%%%%%%%%%%

\begin{figure}
    \centering
    \includegraphics[width=\linewidth]{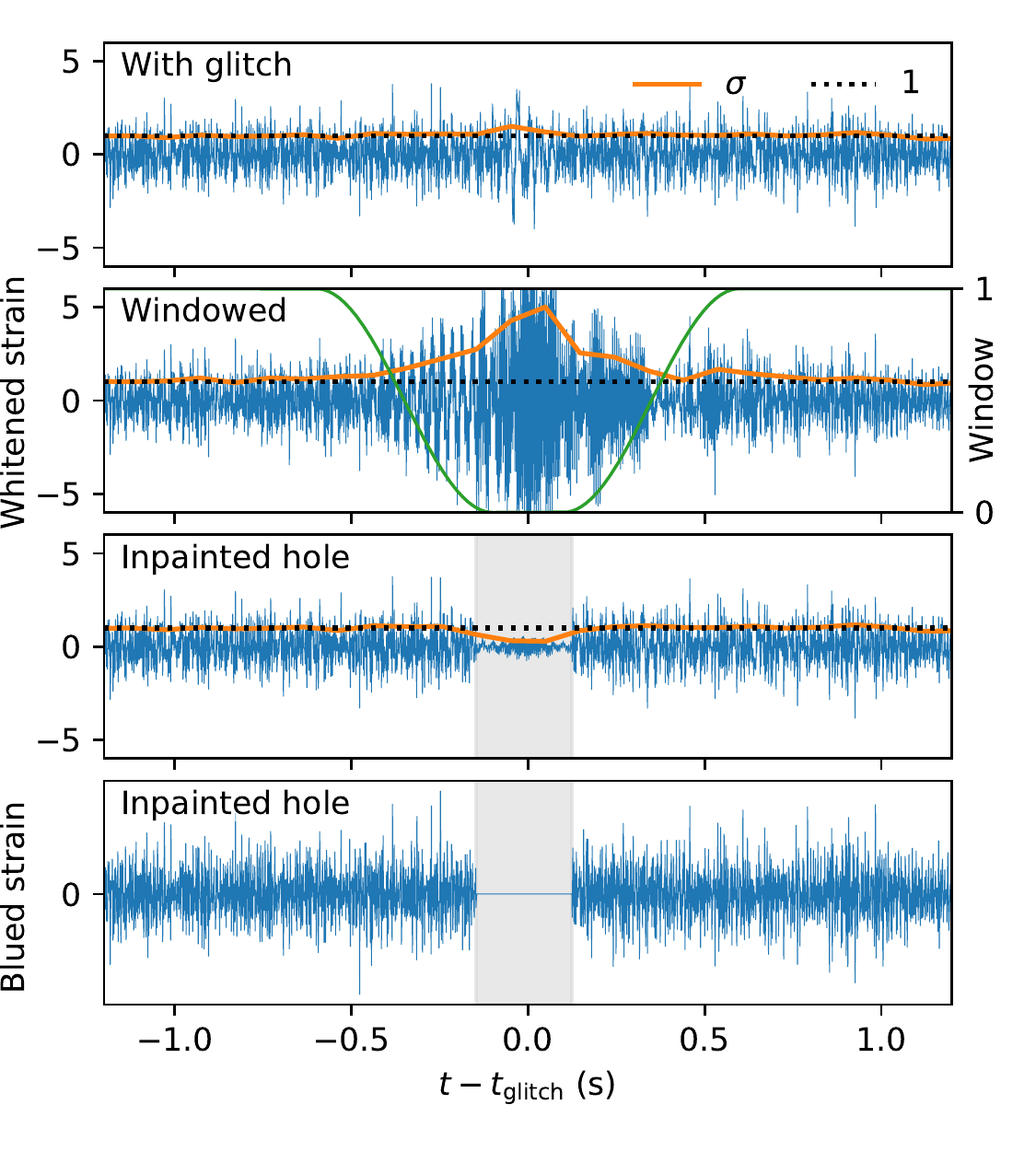}
    \caption{Effect of masking and inpainting glitches. \textit{Top panel:} A segment of whitened strain data (in units of the noise standard deviation) that contains a glitch. The orange curve tracks the standard deviation $\sigma$ calculated from a running window of 100 samples, and is typically close to unity as expected for whitened data. \textit{Second panel:} Gating the glitch with an upside-down Tukey window (green) and then whitening generates artifacts in the whitened data, even outside the Tukey window. For example, $\sigma$ stays above 1.1 for approximately \SI{2}{\second} to each side of the glitch. \textit{ Third panel:} The inpainted whitened data has unit variance outside the hole (shaded). \textit{Bottom panel:} After inpainting, the ``blued" strain is identically zero inside the hole, so overlaps with templates do not depend on the waveform information from inside the hole.
    Figure previewed already in the pipeline description paper, Ref.~\cite{pipelinepaper}.}
    \label{fig:hole_filling}
\end{figure}

\begin{figure}
    \centering
    \includegraphics[width=\linewidth]{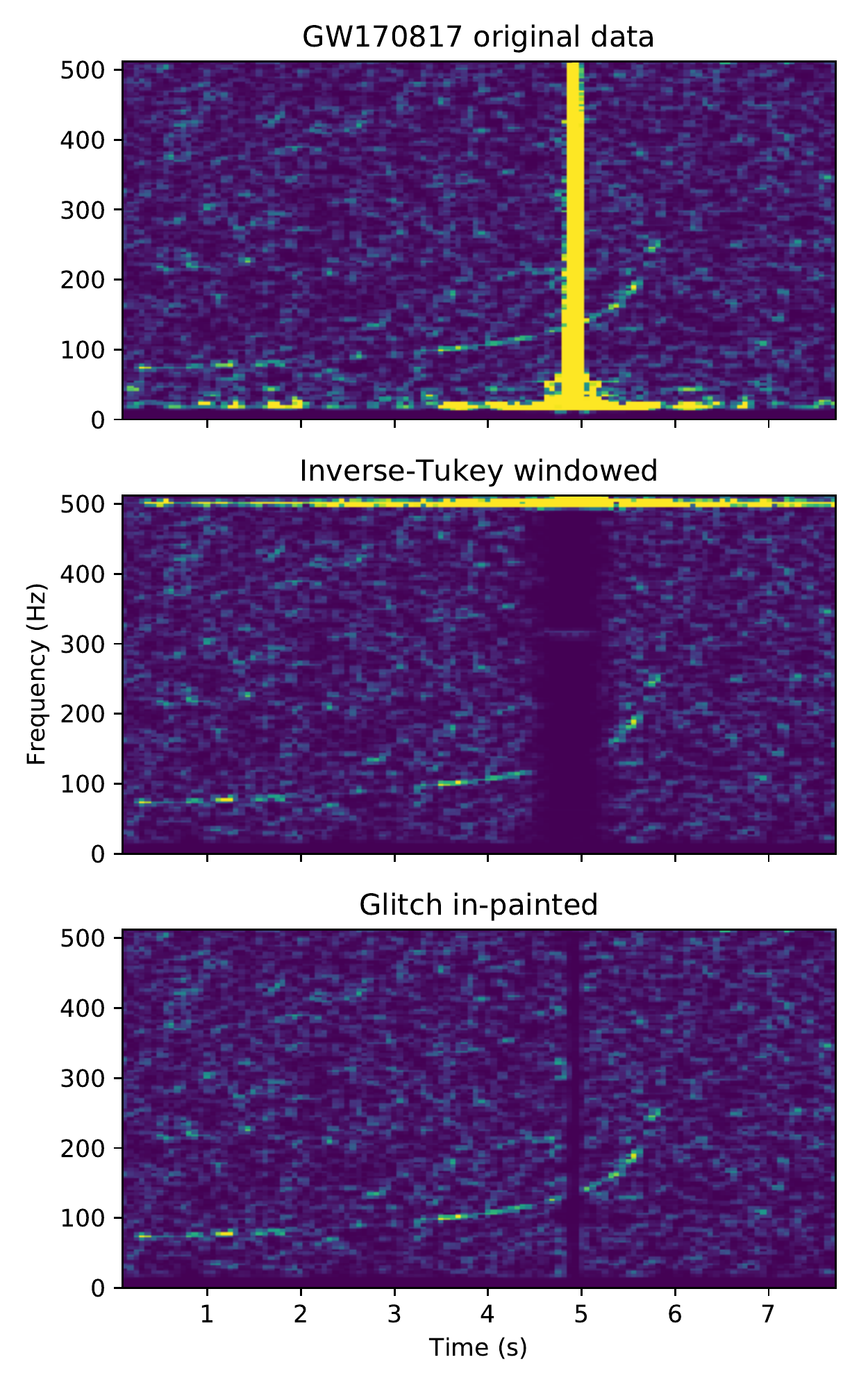}
    \caption{Demonstration of inpainting in the segment of data containing GW170817.\textit{ Upper panel:} whitened spectrogram of the original Livingston strain few seconds around the merger of GW170817.\textit{ Middle panel:} glitch gated away with an inverse Tukey window, with a timescale of \SI{1.2}{\second} and $\alpha=5/6$. Note the leakage from the spectral lines (around \SI{500}{\hertz}) is strongly affecting the data \SI{5}{\second} to either side. A narrower window would leave a more prominent noise leakage.\textit{ Bottom panel}: \SI{160}{\milli\second} inpainted by the hole-filling method presented in this work.}
    \label{fig:hole_filling170817}
\end{figure}

In almost all data segments of $\gtrsim \SI{100}{\second}$, abrupt disturbances are prevalent. Such disturbances have diverse (and often unknown) physical or instrumental origins, and mathematical models that can be used to accurately characterize them are lacking. The timescale of disturbance ranges from a few milliseconds to a few seconds.
During searches for signals from compact binary coalescence, these disturbances induce candidate triggers that populate the tail of the distribution. Their presence dilutes the significance of genuine astrophysical triggers and degrades the search sensitivity. Removing these bad segments of data is not a trivial task, as simply zeroing them out (usually done using a ``gate'' that smoothly zeroizes the data) might result in leakage of excess power to within tens of seconds around, which is often more harmful to the search effort than the disturbance itself.
This is because spectral lines in the PSD whose inverse widths are longer than the duration of the gate leak power to neighboring seconds and frequency bins. See for example the middle panels of Figs.~\ref{fig:hole_filling} and \ref{fig:hole_filling170817}.

%%%%%%%%%%%%%%%%%%%%%%%%%%%%%%%%%%%%%%%%%%%%%%%%%%%%%%%%%%%%%%%%%%
%%%%%%%%%%%%%%%%%%%%%%%%%%%%%%%%%%%%%%%%%%%%%%%%%%%%%%%%%%%%%%%%%%

%%%%%%%%%%%%%%%%%%%%%%%%%%%%%%%%%%%%%%%%%%%%%%%%%%%%%%%%%%%%%%%%%%
%%%%%%%%%%%%%%%%%%%%%%%%%%%%%%%%%%%%%%%%%%%%%%%%%%%%%%%%%%%%%%%%%%

While very often these disturbance induced triggers are easily dismissed, sometimes careful treatment is necessary, as the time-domain footprint of real GW events (especially for BNS events whose waveforms last very long in band) may fortuitously overlap a disturbance. In that case, accurately determining the significance and estimating the merger parameters can be a complicated task. For example, GW170817 was in coincidence with a large disturbance in the Livingston detector (see Fig.~\ref{fig:hole_filling170817}). The analysis by the LIGO and Virgo Collaboration coherently removed this glitch in their analysis of GW170817~\cite{GW170817, Cornish2015}, but most glitches lack an exact mathematical model. We therefore foresee that the analysis of future BNS detections, especially as the sensitivity at low frequencies improves, will necessitate a treatment independent of any exact glitch model. 
In this section, we derive a method to exactly remove bad data segments, ensuring that the significance and inferred parameters of the event are not influenced by the offending segment of the data.

%%%%%%%%%%%%%%%%%%%%%%%%%%%%%%%%%%%%%%%%%%%%%%%%%%%%%%%%%%%%%%%%%%
\subsection{Derivation}
%%%%%%%%%%%%%%%%%%%%%%%%%%%%%%%%%%%%%%%%%%%%%%%%%%%%%%%%%%%%%%%%%%

When computing overlaps one often assumes that the noise covariance is diagonal in the frequency domain and writes:
\begin{align}
    z_w = \sum_{f}{\frac{h^*(f)\,d(f)}{S_w(f)}}.
\end{align}
In practice, the data contain bad seconds that we need to mask out. Let us consider a data series of $N_{d}$ samples and denote it as a vector with components $d_i$. We will denote the $N_d \times N_d$ covariance matrix of the noise as $C_{ij}$, which is diagonal in the frequency domain. Adopting the notation of linear algebra, the overlap can be cast into the matrix form:
\begin{equation}
    z_w=C_{ij}^{-1}\,h_i^*\,d_j=h^{\dagger}\,C^{-1}\,d,
\end{equation}
where $h_i$ are the components of the template waveform and similarly for $d_i$. 

In the presence of loud disturbances, we want the computed overlap to be independent of the behavior of the template waveform within the bad seconds. Let us assume that there is a list of samples in the time domain of length $N_{h}$ to be masked. We denote $u^{(\alpha)}$ a list of $N_h$ vectors such that each of these vectors is zero everywhere except at one of the samples to be masked.  We can define the matrix $A$ of size $N_d\times N_h$:
\begin{equation}
    A_{i,\alpha}=u^{(\alpha)}_i,
\end{equation}
and the matrix $M$ of size $N_h\times N_h$ as:
\begin{equation}
    \begin{split}
        M_{\alpha,\beta} &= C_{ij}^{-1}\,A_{i,\alpha}\, A_{j,\beta} \\
        &= C_{ij}^{-1}\,u^{(\alpha)}_i\,u^{(\beta)}_j \\
        &= A^T\,C^{-1}\,A.
    \end{split}
\end{equation}
We are using the convention that Greek indices run over $1, \ldots, N_h$ and roman indices run over $1, \ldots, N_d$.
We will now define the inpainting filter:
\begin{equation}
    F = 1 - A\,M^{-1}\,A^{T}\,C^{-1}
\end{equation}
and compute the scores with the inpainted data:
\begin{equation}
    \tilde z_w=h^{\dagger}\,C^{-1}\,F\,d. 
\end{equation}
The presence of the hole changes the normalization of the template, so we renormalize $\tilde z$ and compute:
\begin{equation}
    z_w = \bigg(\frac{h^{\dagger}\,C^{-1}\,h}{h^{\dagger}\,C^{-1}\,F\,h}\bigg)^{1/2}\,\tilde z_w.
\end{equation}
Note that this normalization is time and template dependent, which makes it computationally intractable to compute exactly everywhere. In the next subsection, we show how to approximate it using fast Fourier transform (FFT), via the stationary phase approximation.

The inpainting filter $F$ has several desirable properties. First the score does not depend on the values of the template inside the hole because:
\begin{equation}\label{Eq:ZeroResponseInsideHole}
  u^{(\alpha)T}\,C^{-1}\,F\,d = 0,  
\end{equation}
for any $d$. The inpainting filter also satisfies:
\begin{align}
    F^2 &= F \\
    F^T\,C^{-1}\,F &= C^{-1}\,F = F^T\,C^{-1}.
\end{align}
Note that the filter $F$ accomplishes its task by filling the hole with appropriate values. Values outside the holes are left untouched. The term $A\,M^{-1}\,A^T\,C^{-1}$ first weights the data by the inverse of the covariance. Then takes the values inside the holes (by multiplying by $A^{T}$) and makes some linear combinations (by multiplying them by $M^{-1}$). These values are then put back into the hole by multiplying these linear combinations by $A$. Note that the values outside of the holes are left untouched. 

This hole filling procedure ensures that the scores do not depend on what the template does inside the hole. The only requirement is that the covariance matrix used to compute the scores ($C^{-1}$) is the same as the one used to build the matrix $M$.

We can obtain the same inpainting filter by considering a related problem. Suppose that the strain data $d$ is the sum of a  Gaussian random field $h$ with covariance matrix $C$ plus some additional source of noise $n$ with covariance matrix $N$, $d=h+n$. Then the probability of $h$ given the data $d$, $P(h|d)$, is: 
\begin{equation}
    P(h|d) \propto e^{-\frac{1}{2}\,(d-h)^\dagger\,N^{-1}\,(d-h) }\  e^{-\frac{1}{2}\,h^\dagger\,C^{-1}\,h }.
\end{equation}
In the limit where the additional noise is zero outside the holes and infinite inside, the maximum of this function is given by $h = F\,d$. 

Another equivalent formulation is that one is trying to find the maximum of 
\begin{equation}
    \chi^2=\frac{1}{2}\,h^\dagger\,C^{-1}\,h,
\end{equation}
with $h=d+\sum_\alpha\,a_\alpha\,u^{(\alpha)}$ for any $a_\alpha$. That is, find an $h$ that equals the data outside the holes but can take any value inside. This is achieved by taking $h=F\,d$.

%%%%%%%%%%%%%%%%%%%%%%%%%%%%%%%%%%%%%%%%%%%%%%%%%%%%%%%%%%%%%%%%%%
\subsection{Practical issues with inpainting}
%%%%%%%%%%%%%%%%%%%%%%%%%%%%%%%%%%%%%%%%%%%%%%%%%%%%%%%%%%%%%%%%%%

%%%%%%%%%%%%%%%%%%%%%%%%%%%%%%%%%%%%%%%%%%%%%%%%%%%%%%%%%%%%%%%%%%
\subsubsection*{Correcting the variance}
%%%%%%%%%%%%%%%%%%%%%%%%%%%%%%%%%%%%%%%%%%%%%%%%%%%%%%%%%%%%%%%%%%

In order to assess the significance of overlaps computed on hole-filled data, we need to renormalize them using 
\begin{align}
    \lambda_{\rm hole} \equiv \frac{h^{\dagger}\,C^{-1}\,F\,h}{h^{\dagger}\,C^{-1}\,h}.
\end{align}
Under the stationary phase approximation, this coefficient could be understood as the fraction of SNR$^2$ that remains after the segment with bad data has been removed, i.e.,
\begin{align}
    \lambda_{\rm hole}(t_0,h) \approx \frac{\sum_{t\notin {\rm hole}}{|h_{\rm w}(t-t_0)|^2}}{\sum_t {|h_{\rm w}(t)|^2}},
\end{align}
where $t_0$ denotes the merger time and $h_{\rm w}$ is the whitened waveform. Computing this for all times is then done via:
\begin{align}
    \lambda_{\rm hole}(t_0,h) \approx \frac{\left(|\rev{h_{\rm w}}|^2 \conv \mathbbm{1}_{\rm valid}\right)(t_0)}{\sum_t {|h_{\rm w}(t)|^2}},
\end{align}
where $\mathbbm{1}_{\rm valid}$ is one whenever the data is valid and zero otherwise. The convolution operation is computed via FFT using the convolution theorem. 

In principle, the hole correction depends on the merger phase, but under the stationary phase approximation, and the approximation that the waveform would complete many cycles inside the hole, this correction would be the same for waveforms with different phases, and could be computed the same way for complex waveforms.
The stationary phase approximation does not apply to short waveforms (say waveforms that are shorter than a few seconds), for these we do not compute the correction and we simply ignore overlaps that are too close to a hole. For short waveforms the correction is also not relevant as the fraction of data that is removed this way is smaller than a few percent of the total coincident time.

%%%%%%%%%%%%%%%%%%%%%%%%%%%%%%%%%%%%%%%%%%%%%%%%%%%%%%%%%%%%%%%%%%
\subsubsection*{Edge holes}
%%%%%%%%%%%%%%%%%%%%%%%%%%%%%%%%%%%%%%%%%%%%%%%%%%%%%%%%%%%%%%%%%%

When data stream begins after a long break, or a long period of bad data, it is desirable to be able to detect events that are only partially inside the valid data. In order to do this optimally, we treat the edges of the data as holes, and fill them using the same formulae.

%%%%%%%%%%%%%%%%%%%%%%%%%%%%%%%%%%%%%%%%%%%%%%%%%%%%%%%%%%%%%%%%%%
\subsubsection*{Reducing the computational complexity of hole filling}
%%%%%%%%%%%%%%%%%%%%%%%%%%%%%%%%%%%%%%%%%%%%%%%%%%%%%%%%%%%%%%%%%%

Since computing $Fd$ requires inverting a matrix of size $N_h\times N_h$, which costs $N_h^3$ operations, it is of importance to find efficient ways of computing those.
Since the whitening filter is of finite size (fixed in our pipeline to \SI{64}{\second}), filling holes that are \SI{64}{\second} apart could be done independently.

Hole filling a contiguous segment could be done faster via Toeplitz linear equation solution. This can be done through solving Eq.~\eqref{Eq:ZeroResponseInsideHole} for the vector $Fd$, using the fact that $C^{-1}$ is Toeplitz, and that the only disturbed values are inside the hole.
Segments of data with inseparable holes that have more than \SI{10}{\second} of hole duration are assumed to be all bad and filled as contiguous segments using the Toeplitz solver.
When the size of the hole is bigger than the size of the blueing filter (i.e., the whitening filter applied twice) we simply treat this as an edge, and fill only the duration of the blueing filter.

%%%%%%%%%%%%%%%%%%%%%%%%%%%%%%%%%%%%%%%%%%%%%%%%%%%%%%%%%%%%%%%%%%
\section{Summary}
%%%%%%%%%%%%%%%%%%%%%%%%%%%%%%%%%%%%%%%%%%%%%%%%%%%%%%%%%%%%%%%%%%

We have presented methods to deal with two kinds of non-Gaussianity in the strain data, one kind is the fast changes of the PSD, and the other is times of bad data containing abrupt noise transients. 

We proposed computing a running estimate for the variance of the overlaps. We call this correction the PSD drift correction and we have shown that it correctly accounts for the first order errors due to having a wrong PSD in calculating the overlaps. We have shown that the error in estimating the overlap variance is of order $2\%$ on \SI{15}{\second} of data, a segment that is short enough so that error due to even faster PSD variations is of similar magnitude. 
Albeit the PSD drift effect in principle has complex frequency structure, the correction could be done via a simple robust running mean estimate for the local variance of the overlaps.

We have shown that applying this correction dramatically reduces the background in single-detector overlap distributions, and because of this the search volume is increased by at least $15\%$. This improvement is more substantial if the detection significance mainly hinges on one detector, as often happens due to the different antenna patterns of the detectors or their respective sensitivities. We point out that finding the physical source of PSD change or  better tracking of the fast changes could further improve the search sensitivity by $2-3\%$.

We have demonstrated a method to inpaint the strain data in unusable data segments. We have presented how this method allows to ignore loud strain artifacts, compute exact significance for genuine GW events that step into these loud artifacts and estimate the exact significance of such candidates. We expect all these methods to be instrumental in GW search in current and future data analysis.

%%%%%%%%%%%%%%%%%%%%%%%%%%%%%%%%%%%%%%%%%%%%%%%%%%%%%%
\section*{Acknowledgments}
%%%%%%%%%%%%%%%%%%%%%%%%%%%%%%%%%%%%%%%%%%%%%%%%%%%%%%

We greatly thank the LIGO Collaboration and the Virgo Collaboration for making the O1 and O2 data publicly accessible and easily usable.

This research has made use of data, software and/or web tools obtained from the Gravitational Wave Open Science Center (https://www.gw-openscience.org), a service of LIGO Laboratory, the LIGO Scientific Collaboration and the Virgo Collaboration. LIGO is funded by the U.S. National Science Foundation. Virgo is funded by the French Centre National de Recherche Scientifique (CNRS), the Italian Istituto Nazionale della Fisica Nucleare (INFN) and the Dutch Nikhef, with contributions by Polish and Hungarian institutes.

BZ acknowledges the support of The Peter Svennilson Membership
fund. LD acknowledges the support from the Raymond and Beverly Sackler Foundation Fund. TV acknowledges support by the Friends of the Institute for Advanced Study.
MZ is supported by NSF grants AST-1409709,  PHY-1521097 and  PHY-1820775 the Canadian Institute for Advanced Research (CIFAR)
program on Gravity and the Extreme Universe and the Simons Foundation Modern Inflationary Cosmology initiative.

\onecolumngrid
\appendix

%%%%%%%%%%%%%%%%%%%%%%%%%%%%%%%%%%%%%%%%%%%%%%%%%%%%%%%%%%%%%%%%%%
\section{Statistical properties of non-stationary noise}
\label{ap:varyingnoise}
%%%%%%%%%%%%%%%%%%%%%%%%%%%%%%%%%%%%%%%%%%%%%%%%%%%%%%%%%%%%%%%%%%

\subsection{Covariance between Fourier components}
\label{ap:varyingnoisecov}

In this section, we derive the correction to the definition of the discrete PSD, i.e., Eq.~\eqref{eq:psdmeanvar} in the presence of slow variations in the properties of the noise. The starting point are Eqs.~\eqref{eq:nsautocorr} and \eqref{eq:nspsd0}, which define the behavior of the noise over short and long timescales.

As in Section \ref{sec:PSDEstimation}, we work with DFT coefficients defined over a segment of data of length $N$. We are interested in the correlation between the DFT coefficients, which we expand as 
\begin{align}
  \llangle \tilde{d} \lt f_m \rt \left[ \tilde{d} \lt f_{m^\pr} \rt \right]^\ast \rrangle & = \sum_{a, b} \llangle d(a \Delta t) d(b \Delta t) \rrangle e^{2 \pi i \lt f_{m^\pr} b - f_{m} a \rt \Delta t} \label{eq:covmatdft}.
\end{align}
Let us first work through the stationary case, where the expectation value on the right-hand side equals
\begin{align}
  \llangle \tilde{d} \lt f_m \rt \left[ \tilde{d} \lt f_{m^\pr} \rt \right]^\ast \rrangle & = \sum_{a, b} C_{\rm n}\left[ (b - a) \Delta t\right] e^{2 \pi i \lt f_{m^\pr} b - f_{m} a \rt \Delta t}.
\end{align}
Substituting the inverse of Eq.~\eqref{eq:psddef}, we have
\begin{align}
  \llangle \tilde{d} \lt f_m \rt \left[ \tilde{d} \lt f_{m^\pr} \rt \right]^\ast \rrangle & = \sum_{a, b} \intr {\rm d}f S_{{\rm n}, 2}(f) e^{2\pi i f (b - a) \Delta t} e^{2 \pi i \lt f_{m^\pr} b - f_{m} a \rt \Delta t} \\
  & = \intr {\rm d}f S_{{\rm n}, 2}(f) \sum_{b = 0}^N e^{2\pi i \lt f + f_{m^\pr} \rt b \Delta t} \sum_{a = 0}^N e^{- 2\pi i \lt f + f_m \rt a \Delta t}.
\end{align}
In the above equations, indices $a$ and $b$ run from $0$ to $N-1$ in steps of 1 (as in the definition of the DFT in Footnote \ref{fn:dft}). Without loss of generality, we can shift the time-axis so that they run from $-N/2 + 1$ to $N/2$ instead. Evaluating the sums on the right-hand side of the above equation, we get
\begin{align}
  \llangle \tilde{d} \lt f_m \rt \left[ \tilde{d} \lt f_{m^\pr} \rt \right]^\ast \rrangle & = e^{\pi i \lt f_{m^\pr} - f_m \rt \Delta t} \intr {\rm d}f S_{{\rm n}, 2}(f) W_N \lt f; f_m \rt W_N \lt f; f_{m^\pr} \rt, 
\end{align}
where $W_N \lt f; f_m \rt$ is the following window function:
\begin{align}
  W_N \lt f; f_m \rt &= \frac{\sin{[\pi \lt f + f_m \rt N \Delta t]}}{\sin{[\pi \lt f + f_m \rt \Delta t]}}. \label{eq:window}
\end{align}
As a function of frequency $f$, $W_N \lt f; f_m \rt$ has a series of peaks of amplitude $N$ and alternating signs (for even $N$), and width $\sim 1/(N\Delta t)$, that pick out frequencies where the sines in the numerator and denominator vanish. 
The peak-frequencies are separated by the sampling frequency, so if we bandpassed the data before sampling, we can restrict to the primary frequency interval between $-f_{\rm s}/2$ and $f_{\rm s}/2$ (where $f_{\rm s} = 1/\Delta t$ is the sampling frequency).
If the PSD behaves smoothly on frequency scales of $\sim 1/(N\Delta t)$, $W_N \lt f; f_m \rt$ behaves like a delta-function selecting the appropriate frequency in the integrand, and we have
\begin{align}
  \llangle \tilde{d} \lt f_m \rt \left[ \tilde{d} \lt f_{m^\pr} \rt \right]^\ast \rrangle \approx \frac{N}{\Delta t} S_{{\rm n}, 2}\lt f_m \rt \delta_{m, m^\pr} = \frac{N}{2 \Delta t} S_{\rm n} \lt f_m \rt \delta_{m, m^\pr}. \label{eq:psdmeanvar_app}
\end{align}
Applying the whitening filter, we obtain
\begin{align}
  \llangle \wt{d_{\rm w}} \lt f_m \rt \left[ \wt{d_{\rm w}} \lt f_{m^\pr} \rt \right]^\ast \rrangle & = N \delta_{m, m^\pr}. \label{eq:psdmeanvar_white_app}
\end{align}
Next, we now include the non-stationary terms (the extra terms in Eqs.~\eqref{eq:nsautocorr} and \eqref{eq:nspsd0}) when simplifying Eq.~\eqref{eq:covmatdft}. They lead to an extra term, which we evaluate as follows:
\begin{align}
  \delta \llangle \tilde{d} \lt f_m \rt \left[ \tilde{d} \lt f_{m^\pr} \rt \right]^\ast \rrangle & = \sum_{a, b} \delta C_{\rm n}\left[ (b-a) \Delta t; \frac{a + b}{2} \Delta t \right] e^{2 \pi i \lt f_{m^\pr} b - f_{m} a \rt \Delta t} \\
  & = \sum_{a, b} \intr {\rm d}f \, S_{{\rm n}, 2}(f) \epsilon \left[ f; \frac{a + b}{2} \Delta t \right] e^{2\pi i f (b - a) \Delta t} e^{2 \pi i \lt f_{m^\pr} b - f_{m} a \rt \Delta t} \label{eq:covmatnonstationary0},
\end{align}
where in the second equation, we used the inverse of Eq.~\eqref{eq:nspsd1}. It is convenient to define the variables $A$ and $B$:
\begin{align}
  A & = \frac{a + b}{2} \in \left\{ -\frac{N}{2} + 1, -\frac{N}{2} + \frac{3}{2}, \dots, \frac{N}{2} \right\}, \, {\rm and} \\
  B & = b - a \in \left\{ -L_A,  L_A + 2, \dots, L_A \right\}, \, {\rm where} \\
  L_A & = 
  \begin{cases}
    N - 2 + 2 A, & A \leq \frac{1}{2}, \\
    N - 2 A, & A > \frac{1}{2}
  \end{cases}.
\end{align}
We recast Eq.~\eqref{eq:covmatnonstationary0} in terms of these variables and simplify as follows:
\begin{align}
  \delta \llangle \tilde{d} \lt f_m \rt \left[ \tilde{d} \lt f_{m^\pr} \rt \right]^\ast \rrangle & = \sum_A \sum_{B = -L_A}^{L_A} \intr {\rm d}f \, S_{{\rm n}, 2}(f) \epsilon( f; A \Delta t ) e^{2\pi i f B \Delta t} e^{2 \pi i \left[ f_{m^\pr} \lt A + B/2 \rt - f_{m} \lt A - B/2 \rt \right] \Delta t} \\
  & = \sum_A \intr {\rm d}f \, S_{{\rm n}, 2}(f) \epsilon( f; A \Delta t ) e^{2 \pi i \lt f_{m^\pr} - f_m \rt A \Delta t} \sum_{B = -L_A}^{L_A} e^{2 \pi i \left[ f + \lt f_{m^\pr} + f_m \rt/2 \right] B \Delta t} \\
  & = \sum_A \intr {\rm d}f \, S_{{\rm n}, 2}(f) \epsilon( f; A \Delta t ) e^{2 \pi i \lt f_{m^\pr} - f_m \rt A \Delta t} W_{L_A + 1} \lt 2 f; f_m + f_{m^\pr} \rt.
\end{align}
As earlier, we restrict to the primary frequency interval between $-f_{\rm s}/2$ and $f_{\rm s}/2$, within which the window function picks out two peaks at $f = - \bar{f}$ and $f = \pm f_{\rm s}/2 - \bar{f}$ (here, $\bar{f} = (f_m + f_{m^\pr})/2$, and in the second equation, we pick the sign that brings the right-hand side into the primary interval). 
The peaks have the same (opposite) signs when $L_A$ is even (odd).
The peaks are smeared out when $L_A$ is of order unity, which happens only near the edges of the interval.
We neglect these edge effects and replace the window function with the appropriate delta functions at its peaks:
\begin{align}
  \delta \llangle \tilde{d} \lt f_m \rt \left[ \tilde{d} \lt f_{m^\pr} \rt \right]^\ast \rrangle & \approx \sum_A \intr {\rm d}f \, S_{{\rm n}, 2}(f) \epsilon( f; A \Delta t ) e^{2 \pi i \lt f_{m^\pr} - f_m \rt A \Delta t} \frac{1}{2 \Delta t} \times \notag \\
  & \hspace{50pt} \begin{cases} 
  \left[ \delta\lt f + \bar{f} \rt - \delta\lt f + \bar{f} \pm \frac{f_{\rm s}}{2} \rt \right] & A \in \left\{ -\frac{N}{2} + \frac32, -\frac{N}{2} + \frac52 \dots, \frac{N}{2} - \frac12 \right\}, \\
  \left[ \delta\lt f + \bar{f} \rt + \delta\lt f + \bar{f} \pm \frac{f_{\rm s}}{2} \rt \right] & A \in \left\{ -\frac{N}{2} + 1, -\frac{N}{2} + 2 \dots, \frac{N}{2} \right\}
  \end{cases}.
\end{align}
Let us denote the frequency offset $f_m - f_{m^\pr}$ by $\Delta f$. The contributions of the second term inside the square brackets to the sum approximately cancel (since the parts with alternating signs are approximately equal), and hence we can simplify the sum to
\begin{align}
  \delta \llangle \tilde{d} \lt f_m \rt \left[ \tilde{d} \lt f_{m^\pr} \rt \right]^\ast \rrangle 
  & \approx \frac{1}{2 \Delta t} \sum_A S_{{\rm n}, 2} \lt -\bar{f} \rt \epsilon \lt -\bar{f}; A \Delta t \rt e^{- 2 \pi i \Delta f A \Delta t} \label{eq:spvarpsd0} \\
%  & \approx \frac{ S_2 \lt -\bar{f} \rt }{(\Delta t)^2} \tilde{\epsilon} \lt -\bar{f}; \Delta f \rt 
%  = \frac{ S\lt \bar{f} \rt}{2(\Delta t)^2} \tilde{\epsilon} \lt \bar{f}; \Delta f \rt \label{eq:spvarpsd1},
  & \approx \frac{ S_{\rm n} \lt \bar{f} \rt}{2 \Delta t} \tilde{\epsilon} \lt \bar{f}; \Delta f \rt \label{eq:spvarpsd1},
\end{align}
where $\tilde{\epsilon} \lt \bar{f}; \Delta f \rt$ is the DFT, evaluated at the frequency offset $\Delta f$, of $\epsilon \lt \bar{f}, T = n \Delta t \rt$ sampled at a rate of $f_{\rm s} = 1/(\Delta t)$.\footnote{Note that the sum in Eq.~\eqref{eq:spvarpsd0} is on a finer grid sampled at frequency $2/(\Delta t)$. Equation \eqref{eq:spvarpsd1} holds assuming that $\epsilon$ is bandlimited below $f_{\rm s} = 1/(\Delta t)$, which is true in our case.}

%with samples on the original with respect to the slow timescale $T$, evaluated at the frequency offset $\Delta f$. To obtain the final equation, we approximated the sum in Eq.~\eqref{eq:spvarpsd0} as a Riemann sum for an integral.

Finally, we combine Eqs.~\eqref{eq:psdmeanvar_app} and \eqref{eq:spvarpsd1} and apply the whitening filter to obtain the generalized version of Eq.~\eqref{eq:psdmeanvar_white}:
\begin{align}
    \llangle \wt{d_{\rm w}} \lt f_m \rt \left[ \wt{d_{\rm w}} \lt f_{m^\pr} \rt \right]^\ast \rrangle 
    & \approx N \delta_{m, m^\pr} + \frac{ S_{\rm n} \lt \bar{f} \rt }{ \left[ S_{\rm n} \lt f_m \rt S_{\rm n} \lt f_{m^\pr} \rt \right]^{1/2} } \tilde{\epsilon} \lt \bar{f}; \Delta f \rt.
\end{align}
We assume that (a) the changes to the PSD $S_{\rm n}(f)$ occur on long timescales, i.e., $\Delta f \ll \bar{f}$, and (b) $S_{\rm n}(f)$ is smooth on frequency scales $\sim \Delta f$. Then, we can write
\begin{align}
    \llangle \wt{d_{\rm w}} \lt f_m \rt \left[ \wt{d_{\rm w}} \lt f_{m^\pr} \rt \right]^\ast \rrangle 
    & \approx N \delta_{m, m^\pr} + \tilde{\epsilon} \lt \bar{f}; \Delta f \rt. \label{eq:nspsdmeanvar_app}
\end{align}
In our application, the first assumption above holds very well, while the second assumption fails only in the immediate vicinity of spectral lines.

\subsection{Power spectrum of the variance of the matched filtering scores}
\label{ap:powerspectrumvar}

In this section, we consider the variance of the matched filtering scores $z(t)$ (as defined in Eq.~\eqref{eq:matchedFiltertimeseries}) as a time series, and derive its PSD. 
We empirically define the variance as the output of the estimator:
\begin{align}
  \langle z^2 \rangle (t = n \Delta t) & = \sum_{n^\pr} z^2 \left[ t = (n + n^\pr) \Delta t \right] w(-n^\pr \Delta t),
\end{align}
where $\sum_{n} w(n \Delta t) = 1$. 
The window function $w$ effectively manifests as a multiplicative low-pass filter in the formulae in the rest of this section, and does not change any details below its cutoff frequency. 
For ease of presentation, we will omit the window function and directly use the power, $z^2$, in place of the variance $\langle z^2 \rangle$.

The DFT of the power is
\begin{align}
  \wt{z^2} \lt f_m \rt & = \frac{1}{N} \sum_p \left[ \wt{h_{\rm w}} \lt f_p \rt \right]^\ast \wt{d_{\rm w}} \lt f_p \rt \wt{h_{\rm w}} \lt f_p - f_m \rt \left[ \wt{d_{\rm w}} \lt f_p - f_m \rt \right]^\ast. \label{eq:powerdft}
\end{align}
In the above equation, the convention is that frequencies outside the primary interval of $(-f_{\rm s}/2, f_{\rm s}/2]$ are replaced by the values in the interval that they alias to. We view the power as a random time series, with its own autocorrelation function. The PSD of this random series is defined by:
\begin{align}
  \llangle \wt{z^2} \lt f_m \rt \left[ \wt{z^2} \lt f_{m^\pr} \rt \right]^\ast \rrangle & = \frac{N}{2 \Delta t} S_{z^2} \lt f_m \rt \delta_{m, m^\pr}. \label{eq:psdpowermeanvar}
\end{align}
We substitute Eq.~\eqref{eq:powerdft} and obtain:
\begin{align}
  \llangle \wt{z^2} \lt f_m \rt \left[ \wt{z^2} \lt f_{m^\pr} \rt \right]^\ast \rrangle & = 
  \frac{1}{N^2} \sum_{p, q} \left[ \wt{h_{\rm w}} \lt f_p \rt \right]^\ast \wt{h_{\rm w}} \lt f_p - f_m \rt \wt{h_{\rm w}} \lt f_q \rt \left[ \wt{h_{\rm w}} \lt f_q - f_{m^\pr} \rt \right]^\ast \times \notag \\
  & \hspace{50pt} \llangle \wt{d_{\rm w}} \lt f_p \rt \left[ \wt{d_{\rm w}} \lt f_p - f_m \rt \right]^\ast \left[ \wt{d_{\rm w}} \lt f_q \rt \right]^\ast \wt{d_{\rm w}} \lt f_q - f_{m^\pr} \rt \rrangle. \label{eq:psdpower}
\end{align}
We first work through the case with stationary noise. We use Wick's theorem to simplify the expectation value on the right-hand side as the sum of pairwise products, which we evaluate using Eq.~\eqref{eq:psdmeanvar_white_app}:
\begin{align}
  \llangle \wt{z^2} \lt f_m \rt \left[ \wt{z^2} \lt f_{m^\pr} \rt \right]^\ast \rrangle & = 
  \sum_{p, q} \left[ \wt{h_{\rm w}} \lt f_p \rt \right]^\ast \wt{h_{\rm w}} \lt f_p - f_m \rt \wt{h_{\rm w}} \lt f_q \rt \left[ \wt{h_{\rm w}} \lt f_q - f_{m^\pr} \rt \right]^\ast  \left[ \delta_{m, 0} \delta_{m^\pr, 0} + \delta_{p, q} \delta_{m, m^\pr} + \delta_{p + q, m} \delta_{m, m^\pr} \right]. \label{eq:corr_z2}
\end{align}
We obtain an expression for the PSD of the power, $S_{z^2}$, by substituting Eq.~\eqref{eq:psdpowermeanvar} and simplifying:
\begin{align}
  S_{z^2} \lt f_m \rt & = 
  \frac{ 2 \Delta t }{N} \sum_{p, q} \left[ \wt{h_{\rm w}} \lt f_p \rt \right]^\ast \wt{h_{\rm w}} \lt f_p - f_m \rt \wt{h_{\rm w}} \lt f_q \rt \left[ \wt{h_{\rm w}} \lt f_q - f_m \rt \right]^\ast \left[ \delta_{m, 0} + \delta_{p, q} + \delta_{p + q, m} \right] \\
  & = 2 N \Delta t \, \delta_{m, 0} + \frac{ 4 \Delta t }{N} \sum_{p, q} \llvert \wt{h_{\rm w}} \lt f_p \rt \rrvert^2 \llvert \wt{h_{\rm w}}\lt f_q \rt \rrvert^2 \delta_{p+q, m}. \\
  & = 2 N \Delta t \, \delta_{m, 0} + 4 \Delta t \, \wt{\llvert h_{\rm w} \conv \rev{h_{\rm w}} \rrvert^2} \lt f_m \rt. \label{eq:psdz2}
\end{align}
In the final equation, $h_{\rm w} \conv \rev{h_{\rm w}}$ is the autocorrelation function of the whitened waveform. To simplify the first term on the right-hand side, we used the fact that the SNR in Eq.~\eqref{eq:snr2} is normalized to unity.

In the non-stationary case, the PSD of the power, $S_{z^2}$, also depends on the varying part of the noise PSD, $\epsilon (f, T)$. 
Given a number of frequency-bins, $f_m$ (conjugate to the `fast' timescales), we consider the $\epsilon \lt f_m, T \rt$ as a set of random time-series, varying over long timescales $T$. 
The most general form of the correlations between these time-series is
\begin{align}
  \llangle \tilde{\epsilon} \lt f_m, f_a \rt \left[ \tilde{\epsilon} \lt f_{m^\pr}, f_b \rt \right]^\ast \rrangle & = \frac{N}{2 \Delta t} S_{\epsilon, \lt m, m^\pr \rt} \lt f_a \rt \delta_{a, b}. \label{eq:epsilonpsdgen}
\end{align}
In the above equation, $\tilde{\epsilon} \lt f_m, f_a \rt$ represents the DFT of $\epsilon \lt f_m, T = n \Delta t \rt$ sampled at a rate $f_{\rm s} = 1/(\Delta t)$. 
If we perform the Singular Value Decomposition (SVD) of the matrix $S_{\epsilon, \lt m, m^\pr \rt} \lt f_a \rt$, the basis vectors represent frequency components (with weights given by the coefficients) that vary together.
We assume the simplest possible form of the correlations in Eq.~\eqref{eq:epsilonpsdgen}:
\begin{align}
  \llangle \tilde{\epsilon} \lt f_m, f_a \rt \left[ \tilde{\epsilon} \lt f_{m^\prime}, f_b \rt \right]^\ast \rrangle & = \frac{N}{2 \Delta t} S_{\epsilon} \lt f_a \rt \delta_{a, b}. \label{eq:epsilonpsd_app}
\end{align} 
This implies that all the $\epsilon \lt f_m, T \rt$ vary in step with each other, and with the same amplitude.

We return to Eq.~\eqref{eq:psdpower} to evaluate the PSD of the power, $S_{z^2}$.
We first fix a particular realization of $\epsilon \lt f_m, T \rt$, compute $S_{z^2}$ using the disconnected terms in Eq.~\eqref{eq:nspsdmeanvar_app}, and additionally average over the realizations of $\epsilon$ using Eq.~\eqref{eq:epsilonpsd_app} (we neglect the connected part of the expectation value in Eq.~\eqref{eq:nspsdmeanvar_app}). 
In addition to the terms in Eq.~\eqref{eq:psdz2}, the following terms show up:
%\begin{align}
%  \delta \llangle \wt{z^2}(f_m) \left[ \wt{z^2}(f_{m^\pr}) \right]^\ast \rrangle & = 
%  \frac{1}{2 N \Delta t} \sum_{p, q} \left[ \wt{h_{\rm w}}(f_p) \right]^\ast \wt{h_{\rm w}}(f_p - f_m) \wt{h_{\rm w}}(f_q) \left[ \wt{h_{\rm w}}(f_q - f_{m^\pr}) \right]^\ast \times \notag \\
%  & \hspace{50pt} \left[ S_{\epsilon}\lt f_m \rt + S_{\epsilon}\lt f_p - f_q \rt + S_{\epsilon}\lt f_p + f_q - f_m \rt \right] \delta_{m, m^\pr}
%\end{align}
%The resulting correction to the PSD $S_{z^2}$ is
\begin{equation}
    \begin{split}
      \delta S_{z^2} \lt f_m \rt
      &= \frac{1}{N^2} \sum_{p, q} \left[ \wt{h_{\rm w}} \lt f_p \rt \right]^\ast \wt{h_{\rm w}} \lt f_p - f_m \rt \wt{h_{\rm w}} \lt f_q \rt \left[ \wt{h_{\rm w}} \lt f_q - f_m \rt \right]^\ast \left[ S_{\epsilon}\lt f_m \rt + S_{\epsilon}\lt f_p - f_q \rt + S_{\epsilon}\lt f_p + f_q - f_m \rt \right] \\
      &= \vert \wt{h_{\rm w}^2} \lt f_m \rt \vert^2 \, S_\epsilon \lt f_m \rt \\
      &\quad + \frac{1}{N^2} \sum_{p, q} \left[ \wt{h_{\rm w}} \lt f_p \rt \right]^\ast \wt{h_{\rm w}} \lt f_p - f_m \rt \wt{h_{\rm w}} \lt f_q \rt \left[ \wt{h_{\rm w}} \lt f_q - f_m \rt \right]^\ast \left[ S_{\epsilon}\lt f_p - f_q \rt + S_{\epsilon}\lt f_p + f_q - f_m \rt \right]. \label{eq:pertpsdz20}
    \end{split}
\end{equation}
We can see by a change of variable that both the terms in the summation on the right-hand side of Eq.~\eqref{eq:pertpsdz20} are equal. Carrying this through,
\begin{equation}
    \begin{split}
      \delta S_{z^2} \lt f_m \rt
      &= \vert \wt{h_{\rm w}^2} \lt f_m \rt \vert^2 \, S_\epsilon \lt f_m \rt + \frac{2}{N^2} \sum_{p, q} \left[ \wt{h_{\rm w}} \lt f_p \rt \right]^\ast \wt{h_{\rm w}} \lt f_p - f_m \rt \wt{h_{\rm w}} \lt f_q \rt \left[ \wt{h_{\rm w}} \lt f_q - f_m \rt \right]^\ast S_{\epsilon} \lt f_p - f_q \rt \\
      &= \vert \wt{h_{\rm w}^2} \lt f_m \rt \vert^2 \, S_\epsilon \lt f_m \rt \\
      &\quad + \frac{2}{N^2} \sum_{q, \Delta q \equiv (p - q)} \wt{h_{\rm w}} \lt f_q \rt \left[ \wt{h_{\rm w}} \lt f_q + f_{\Delta q} \rt \right]^\ast \left[ \wt{h_{\rm w}} \lt f_q - f_m \rt \right]^\ast \wt{h_{\rm w}} \lt f_q + f_{\Delta q} - f_m \rt \, S_{\epsilon} \lt f_{\Delta q} \rt \\
      &= \vert \wt{h_{\rm w}^2} \lt f_m \rt \vert^2 \, S_\epsilon \lt f_m \rt + \frac{2}{N} \sum_{\Delta q} \wt{\llvert h_{\rm w} \conv \rev{h_{\rm w} e^{- 2 \pi i f_{\Delta q} t}} \rrvert^2} \lt f_m \rt \, S_{\epsilon} \lt f_{\Delta q} \rt. \label{eq:pertpsdz21}
    \end{split}
\end{equation}
In the second term on the right-hand side, we multiplied the signal with an oscillatory function of time ($e^{ - 2 \pi i f_{\Delta q} t}$) before reversing and convolving it with itself. 

We can simplify the form of Eq.~\eqref{eq:pertpsdz21} using the following approximations: (a) variations in the noise PSD ($\tilde{\epsilon}$) have support at much lower frequencies than the whitened signal $\wt{h_{\rm w}}$ does, and (b) the time-domain whitened waveform is much shorter than the timescales over which $\epsilon(f; T)$ varies.
The first assumption holds very well, since $\epsilon(f; T)$ varies on timescales of tens of seconds (i.e., $\tilde{\epsilon}$ has power at $f \lesssim 0.1$ Hz) while signals accumulate SNR only at $f \gtrsim 20$ Hz; the second assumption does not hold for signals that are longer than a few tens of seconds (such as signals from merging binary neutron stars). 

Under the second approximation above, the factor of $e^{- 2 \pi i f_{\Delta q} t}$ in Eq.~\eqref{eq:pertpsdz21} acts like a constant phase multiplying the signal, which makes no difference. We can pull it out, and interpret the summation as a Riemann sum to obtain
\begin{align}
  \delta S_{z^2} \lt f_m \rt & \approx \llvert \wt{h_{\rm w}^2} \lt f_m \rt \rrvert^2 \, S_\epsilon \lt f_m \rt + 2 \Delta t \wt{\llvert h_{\rm w} \conv \rev{h_{\rm w}} \rrvert^2} \lt f_m \rt \intr {\rm d}f \, S_{\epsilon} \lt f \rt. \, \label{eq:pertpsdz22}
\end{align}
We combine Eqs.~\eqref{eq:psdz2} and \eqref{eq:pertpsdz22} to get our final simplified form of the PSD of the variations in the power of the matched filtering scores:
\begin{align}
  S_{z^2} \lt f_m \rt & \approx 2 N \Delta t \, \delta_{m, 0} + 2 \Delta t \, \wt{\llvert h_{\rm w} \conv \rev{h_{\rm w}} \rrvert^2} \lt f_m \rt \left[ 2 + \intr {\rm d}f \, S_{\epsilon} \lt f \rt \right] + \llvert \wt{h_{\rm w}^2} \lt f_m \rt \rrvert^2 \, S_\epsilon \lt f_m \rt. \label{eq:ns_sz2_app}
\end{align}

%%%%%%%%%%%%%%%%%%%%%%%%%%%%%%%%%%%%%%%%%%%%%%%%%%%%%%%%%%%%%%%%%%
\section{Computation of the measurement error in the PSD drift}
\label{ap:ErrorComputation}
%%%%%%%%%%%%%%%%%%%%%%%%%%%%%%%%%%%%%%%%%%%%%%%%%%%%%%%%%%%%%%%%%%

In this section, we work with whitened waveforms and data, $h_{\rm w}(n \Delta t)$ and $d_{\rm w}(n \Delta t)$ respectively, that are functions of the discrete index $n$, and were whitened using some fiducial filter. 
The auto-correlation function of perfectly white data is
\begin{align}
  \langle d_{\rm w}(t) d_{\rm w}(t + n \Delta t) \rangle & = \delta_{n, 0}, \label{eq:wtdat}
\end{align}
where the lowercase $\delta$ stands for the Kronecker delta. We define the time-series of matched filtering overlaps as
\begin{align}
  z(n \Delta t) & = \sum_{m} h_{\rm w} \lt m \Delta t \rt d_{\rm w} \left[ (n - m) \Delta t \right],
\end{align}
which is equivalent to Eq.~\eqref{eq:matchedFiltertimeseries}. 
We now consider the statistics of these overlaps. 

Let us first assume the whitening was successful, i.e., Eq.~\eqref{eq:wtdat} applies. 
In this case, the autocorrelation function of the score $z(t)$ equals that of the waveform, i.e.,
\begin{align}
  \llangle z \lt t \rt z \lt t + n \Delta t \rt \rrangle & = \left[ h_{\rm w} \conv \rev{h_{\rm w}} \right] \lt n \Delta t \rt
%  \sum_{t^\pr, t^\ppr} h_{\rm w}^*(t^\pr) h_{\rm w}(t^\ppr) \langle d_{\rm w}^*(t - t^\pr)  d_{\rm w}(t + \Delta t - t^\ppr) \rangle \\
%  & = \sum_{t^\pr, t^\ppr} h_{\rm w}^*(t^\pr) h_{\rm w}(t^\ppr) \delta(t^\pr - t^\ppr + \Delta t) \\
%  & = \sum_{t^\pr} h_{\rm w}^*(t^\pr) h_{\rm w}(t^\pr + \Delta t).
\end{align}
In particular, if we normalize the whitened waveform to satisfy $\sum_n \vert h_{\rm w} \lt n \Delta t \rt \vert^2 = 1$, the autocorrelation  function equals unity at zero-lag, i.e., the score has unit variance. For typical waveforms, the autocorrelation function has a very small time-domain width (for the waveform for GW150914, the time-domain half-width, defined as the lag at which the autocorrelation function drops to $0.5$, equals $\sim \SI{1.5}{\milli\second}$; at our fiducial sampling rate of \SI{1024}{\hertz}, $\sim 2.5$ samples are correlated). %In what follows, we will neglect this factor, and assume that neighboring scores are uncorrelated in the white noise case.
%\begin{figure}[htbp]
%\centering
%  \includegraphics[width=12cm]{wf_acorr.pdf}
%\caption{Time-domain autocorrelation function of a whitened waveform (we use the central waveform of bank \texttt{BBH (3,0)})}
%\label{fig:wf_acorr}
%\end{figure}

If we fail to whiten the data perfectly, the scores $z(t)$ are still normally distributed, but with a different variance. 
Formally, the new auto-correlation function is
\begin{align}\label{eq:variance}
  V \lt n \Delta t \rt = \llangle z\lt t \rt z \lt t + n \Delta t \rt \rrangle = \sum_{a, b} h_{\rm w} \lt a \Delta t \rt h_{\rm w} \lt b \Delta t \rt \llangle d_{\rm w}^\ast \lt t - a \Delta t \rt  d_{\rm w} \left[ t + (n - b) \Delta t \right] \rrangle. %\\
  %  & = \sum_{n^\pr, n^\ppr} w^*(n^\pr) w(n^\ppr) C(n^\pr - n^\ppr + \Delta n),
\end{align}
We can write the convolutions in Eq.~\eqref{eq:variance} in the Fourier domain.
% \begin{align}
%     \sum_{t^\pr} h_{\rm w}(t^\pr) d_{\rm w}(t - t^\pr)
%     &= \frac{1}{N^2} \sum_{t^\pr} \sum_{p, q} \tilde{h_{\rm w}}(p) \tilde{d_{\rm w}}(q) e^{2\pi i p t^\pr/N} e^{2\pi i q (t - t^\pr)/N} \\ \nonumber
%     &= \frac{1}{N^2} \sum_{p, q}  \sum_{t^\pr} \tilde{h_{\rm w}}(p) \tilde{d_{\rm w}}(q) e^{2\pi i q t/N} e^{2\pi i (p - q) t^\pr/N} \\
%     &= \frac{1}{N^2} \sum_{p, q} \tilde{h_{\rm w}}(p) \tilde{d_{\rm w}}(q) e^{2\pi i q t/N} N \delta(p -q) \\ \nonumber & = \frac{1}{N} \sum_{p} \tilde{h_{\rm w}}(p) \tilde{d_{\rm w}}(p) e^{2\pi i p t/N}
% \end{align}
% where $N$ is an index interval that is large enough to encompass the support of the waveform, and symbols decorated with tildes denote the discrete Fourier transform (DFT) of a time series (with the \texttt{numpy} convention). The autocorrelation function of Eq.~\eqref{eq:variance} is 
\begin{equation}
\begin{split}
  V \lt n \Delta t \rt 
%   &= \sum_{t^\pr, t^\ppr} h_{\rm w}^*(t^\pr) h_{\rm w}(t^\ppr) \langle d_{\rm w}^*(t - t^\pr)  d_{\rm w}(t + \Delta t - t^\ppr) \rangle \\
  &= \frac{1}{N_{\rm a}^2} \sum_{p, q} \left[ \wt{h_{\rm w}} \lt f_p \rt \right]^\ast \wt{h_{\rm w}} \lt f_q \rt \llangle \left[ \wt{d_{\rm w}} \lt f_p \rt \right]^\ast \wt{d_{\rm w}} \lt f_q \rt \rrangle e^{2\pi i \lt f_q - f_p \rt t} e^{2\pi i f_q \Delta t}, \label{eq:variancefourier} 
\end{split}
\end{equation}
where $N_{\rm a}$ is the number of samples used to compute the PSD drift correction. Instead of the general approach we followed in Section \ref{sec:WrongPSD}, we use a segment of data that is shorter than the typical timescales over which the PSD varies, but assume that we used the wrong local PSD. 
In this case, the incorrectly whitened the data, $d_{\rm w}$, is locally described by a stationary and real-valued Gaussian random variable:
\begin{align}
  \llangle \left[ \wt{d_{\rm w}} \lt f_p \rt \right]^\ast \wt{d_{\rm w}} \lt f_q \rt \rrangle & = \frac{N_{\rm a}}{2 \Delta t} S_{d_{\rm w}} \lt f_p \rt \delta_{p, q},
\end{align}
where $S_d$ is the two-sided power spectral density of the process. 
Substituting in Eq.~\eqref{eq:variancefourier} gives us
\begin{align}
  V \lt n \Delta t \rt & = \frac{1}{N_{\rm a} \Delta t} \sum_{p} \frac{1}{2} \llvert \wt{h_{\rm w}} \lt f_p \rt \rrvert^2 S_{d_{\rm w}} \lt f_p \rt e^{2\pi i f_p n \Delta t}. \label{eq:variancepsd}
\end{align}
In the case where $d_{\rm w}$ is white noise, $S_{d_{\rm w}} = 2 \Delta t$, and using Parseval's identity, the zero-lag auto-correlation function of Eq.~\eqref{eq:variancepsd} equals the normalization factor $\sum_n \vert h_{\rm w}(n) \vert^2 = 1$. 

In practice, we use the estimator
\begin{align}
  \mathcal{V} = \frac{1}{N_{\rm a}^2} \sum_p \llvert \wt{h_{\rm w}} \lt f_p \rt \rrvert^2 \llvert \wt{d_{\rm w}} \lt f_p \rt \rrvert^2 = \frac{1}{N_{\rm a}} \sum_{n} z^2 \lt n \Delta t \rt \label{eq:estimator}
\end{align}
The variance of this estimator is 
\begin{equation}
\begin{split}
  \sigma^2_{\mathcal{V}} & = \langle \mathcal{V}^2 \rangle - \langle \mathcal{V} \rangle^2 \\
  & = \frac{1}{N_{\rm a}^4} \sum_{p, q} \llvert \wt{h_{\rm w}} \lt f_p \rt \rrvert^2 \llvert \wt{h_{\rm w}} \lt f_q \rt \rrvert^2 \left[ \llangle \left[ \wt{d_{\rm w}} \lt f_p \rt \right]^\ast \wt{d_{\rm w}} \lt f_p \rt \left[ \wt{d_{\rm w}}\lt f_q \rt \right]^\ast \wt{d_{\rm w}} \lt f_q \rt \rrangle - \llangle \left[ \wt{d_{\rm w}}\lt f_p \rt \right]^\ast \wt{d_{\rm w}} \lt f_p \rt \rrangle \llangle \left[ \wt{d_{\rm w}}\lt f_q \rt \right]^\ast \wt{d_{\rm w}}\lt f_q \rt \rrangle \right] \\
  & = \frac{1}{N_{\rm a}^4} \sum_{p, q} \llvert \wt{h_{\rm w}} \lt f_p \rt \rrvert^2 \llvert \wt{h_{\rm w}} \lt f_q \rt \rrvert^2 \left[ \llangle \left[ \wt{d_{\rm w}} \lt f_p \rt \right]^\ast \left[ \wt{d_{\rm w}} \lt f_q \rt \right]^\ast \rrangle \llangle \wt{d_{\rm w}} \lt f_p \rt \wt{d_{\rm w}} \lt f_q \rt \rrangle + \llangle \left[ \wt{d_{\rm w}} \lt p \rt \right]^\ast \wt{d_{\rm w}} \lt q \rt \rrangle \llangle \wt{d_{\rm w}} \lt p \rt \left[ \wt{d_{\rm w}} \lt q \rt \right]^\ast \rrangle \right] \\
%   & = \frac{1}{N^4} \sum_{p, q} \vert \tilde{h_{\rm w}}(p) \vert^2 \vert \tilde{h_{\rm w}}(q) \vert^2 \left[ \langle \tilde{d_{\rm w}}^*(p) \tilde{d_{\rm w}}(-q) \rangle \langle \tilde{d_{\rm w}}^*(-p) \tilde{d_{\rm w}}(q) \rangle + \langle \tilde{d_{\rm w}}^*(p) \tilde{d_{\rm w}}(q) \rangle \langle \tilde{d_{\rm w}}(p) \tilde{d_{\rm w}}^*(q) \rangle \right] \\
  & = \frac{1}{N_{\rm a}^4} \sum_{p, q} \llvert \wt{h_{\rm w}} \lt f_p \rt \rrvert^2 \llvert \wt{h_{\rm w}} \lt f_q \rt \rrvert^2 \left( \frac{N}{2 \Delta t} \right)^2 S_{d_{\rm w}}^2 \lt f_p \rt \left[ \delta_{p, -q} + \delta_{p, q} \right] \\
  & = \frac{1}{2 \lt N_{\rm a} \Delta t \rt^2} \sum_{p} \llvert \wt{h_{\rm w}} \lt f_p \rt \rrvert^4 S_{d_{\rm w}}^2 \lt f_p \rt.
\end{split}
\end{equation}
We used Wick's theorem in going from the second to the third line, and the real-valued nature of the template and the data in the subsequent lines. If the data $d_{\rm w}$ were white, as earlier, we have $S_{d_{\rm w}} = 2 \Delta t$. Substituting this, we get
\begin{align}
  \sigma^2_{\mathcal{V}} \biggr\vert_{{\rm white} \, d_{\rm w}} & = \frac{2}{N_{\rm a}^2} \sum_{p} \llvert \wt{h_{\rm w}} \lt f_p \rt \rrvert^4 = \frac{4}{N_{\rm a}^2} \sum_{p>0} \llvert \wt{h_{\rm w}} \lt f_p \rt \rrvert^4 . \label{eq:variancetol}
\end{align}
In the above formula, note that the FFT of the whitened template is performed on the interval of length $N_{\rm a}$. In practice, we want to solve for $N_{\rm a}$ given a particular tolerance requirement. Suppose we have the whitened template on a frequency grid conjugate to a fiducial window of length $N_0$. From the connection between the DFT and the continuous Fourier transform, we can scale the sum in Eq.~\eqref{eq:variancetol} according to the size of the grid. 
\begin{align}
  \sigma^2_{\mathcal{V}} \biggr\vert_{{\rm white} \, d_{\rm w}} & \approx \frac{4}{N_{\rm a} N_0} \sum_{p_{N_0} > 0} \llvert \wt{h_{\rm w}} \lt f_{p_{N_0}} \rt \rrvert^4\,. \label{eq:sigmav}
%   \sigma_{\mathcal{V}} \biggr\vert_{{\rm white} \, d_{\rm w}} & \approx \bigg( \frac{4}{N N_0} \sum_{p_{N_0} > 0} \lvert \tilde{h_{\rm w}}(p_{N_0}) \vert^4 \bigg)^{1/2}\,. \label{eq:sigmav}
\end{align}
The formula above suggests that to measure the $(\lambda_w)^2 = \langle \mathcal{V} \rangle = V$ to two percent relative error, we need $\mathcal O(\num{15000})$ samples, i.e., $\sim\SI{15}{\second}$ for a typical waveform.

\twocolumngrid
\bibliography{mainbibfile}% Produces the bibliography via BibTeX.

\end{document}